# Planetary Spectrum Generator: an accurate online radiative transfer suite for atmospheres, comets, small bodies and exoplanets

Villanueva, G.L.[1]; Smith, M. D.[1]; Protopapa, S.[2]; Faggi, S.[1,3]; Mandell, A. M.[1]

(1) *NASA Goddard Space Flight Center, Greenbelt, MD, 20771, USA*

(2) *Department of Space Studies, Southwest Research Institute, Boulder, CO 80302, USA*

(3) *Universities Space Research Association (USRA), Columbia, MD 21046, USA*

## Abstract

We have developed an online radiative-transfer suite (https://psg.gsfc.nasa.gov) applicable to a broad range of planetary objects (e.g., planets, moons, comets, asteroids, TNOs, KBOs, exoplanets). The Planetary Spectrum Generator (PSG) can synthesize planetary spectra (atmospheres and surfaces) for a broad range of wavelengths (UV/Vis/near-IR/IR/far-IR/THz/sub-mm/Radio) from any observatory (e.g., JWST, ALMA, Keck, SOFIA), any orbiter (e.g., ExoMars, Juno), or any lander (e.g., MSL). This is achieved by combining several state-of-the-art radiative transfer models, spectroscopic databases and planetary databases (i.e., climatological and orbital). PSG has a 3D (three-dimensional) orbital calculator for most bodies in the solar system, and all confirmed exoplanets, while the radiative-transfer models can ingest billions of spectral signatures for hundreds of species from several spectroscopic repositories. It integrates the latest radiative-transfer and scattering methods in order to compute high resolution spectra via line-by-line calculations, and utilizes the efficient correlated-k method at moderate resolutions, while for computing cometary spectra, PSG handles non-LTE and LTE excitation processes. PSG includes a realistic noise calculator that integrates several telescope / instrument configurations (e.g., interferometry, coronagraphs) and detector technologies (e.g., CCD, heterodyne detectors, bolometers). Such an integration of advanced spectroscopic methods into an online tool can greatly serve the planetary community, ultimately enabling the retrieval of planetary parameters from remote sensing data, efficient mission planning strategies, interpretation of current and future planetary data, calibration of spectroscopic data, and development of new instrument/spacecraft concepts.

## 1.0    Introduction

Instruments and tools to characterize planetary atmospheres and surfaces have reached an unprecedented level of sophistication and maturity, opening new windows in the exploration of our solar system and beyond. High-resolution infrared spectrometers with broad spectral coverage at ground-based observatories (e.g., Keck, IRTF, VLT) and arrays of radio telescopes with state of the art receivers (e.g., ALMA) now permit the exploration of the kinematics, composition and thermal structure of a broad range of planetary sources with unprecedented precision. These, combined with the advent of comprehensive spectroscopic databases containing billions of lines and accurate reflectance spectra, robust radiative transfer models, and unprecedented available computational power, are transforming the way we investigate planetary objects.

For instance, remote sensing of atmospheric isotopic ratios is now more accessible than ever, and it has recently permitted us to infer that Mars lost an ocean's worth of water [1], and to further understand the role of comets in delivering the water to Earth's oceans [2,3]. Similarly, thanks to modern spectroscopy the spectacular infrared maps derived by the New Horizons mission [4,5] can now be quantitatively interpreted to establish the composition of Pluto's surface with great accuracy and precision [6]. At longer wavelengths, the ALMA array is revolutionizing the characterization of planetary atmospheres by enabling high-spatial resolutions and unparalleled sensitivities, as demonstrated by the surprising maps of HNC and $HC_3N$ obtained of Titan [7] and the observations of Pluto's atmosphere [8].

Interpreting this wealth of planetary data has only been possible thanks to decades of meticulous work by hundreds of laboratory spectroscopists (e.g., [9–11]) and radiative transfer modelers (e.g., [12,13]). This also means that many of these tools have reached a high level of maturity, and also of fractionation. For instance, there are two complementary molecular infrared spectroscopic databases (HITRAN [14], GEISA [15]), several radio molecular databases (e.g., JPL [16], CDMS [17], Spatalogue), and numerous reflectance libraries (e.g., ASTER, RELAB, SpecLib). In many cases, these databases provide conflicting information, and each use particular/unique file formats, that restrict their portability and validation across different radiative transfer applications.

Particularly challenging is the existence of numerous radiative transfer packages for ingesting these databases. Typically, these packages require complex installation and compilation procedures for them to operate, and they are particularly restrictive in operational scope (e.g., planet, type of atmosphere/surface) and wavelength (e.g., spectral database, file formats). Even at the internet encyclopedia (Wikipedia), there is an entry for "Atmospheric radiative transfer codes", listing dozens of packages and their capabilities (wavelength range, geometry, scattering, polarization, accessibility/licensing, etc.). There have been several attempts to quantify the differences between different packages (e.g., [18]), and a commercial internet facility (www.spectralcalc.com) implements a small subset relevant to Earth/Mars science. However, due to a lack of portability in wavelength and geometry, there is currently no "gold-standard" or benchmark across a broad range of observing conditions.

A successful example of the consolidation of multiple data sources occurred in the field of planetary ephemeris, with the creation of the JPL/Horizons online tool in 1996. This tool, which is now widely used by the planetary community, has become the primary repository and tool to compute orbital parameters, thanks to its accuracy, ease of use and online presence. The tool ingests astrometric information collected from a multitude of sources and employs a robust orbital calculator to produce user-friendly lists of orbital parameters that can be used for several planetary applications. A similar successful story is that of the Mars Climate Database [19], which has become the main repository for weather forecasts and circulation models of Mars. The key component that made these tools highly successful was the user-friendly online portal, permitting the public to obtain accurate planetary information without having to compile and install a multitude of software packages (and to learn how to properly operate them).

In an attempt to materialize a similar solution for planetary spectroscopy, we developed the Planetary Spectrum Generator (PSG), an easy-of-use online tool that can ingest a broad range of spectroscopic information while employing accurate models to synthesize planetary fluxes (see figure 1). The applicability of the tool is extensive, from planning observations (e.g., observing proposals, mission planning), to interpreting ground-based and spacecraft planetary data, to developing new instrument/telescope concepts, to calibrating spectroscopic mission data. Beyond the online access, the tool was conceptualized to be accurate and extremely flexible, in order to provide a self-consistent and comprehensive solution for such a broad range of problems. Most importantly, the tool incorporates a set of modern and state-of-the-art radiative transfer packages, ensuring realistic simulations and in spite of the complexity of these simulations, the tool also guarantees short run times and highly efficient computations.

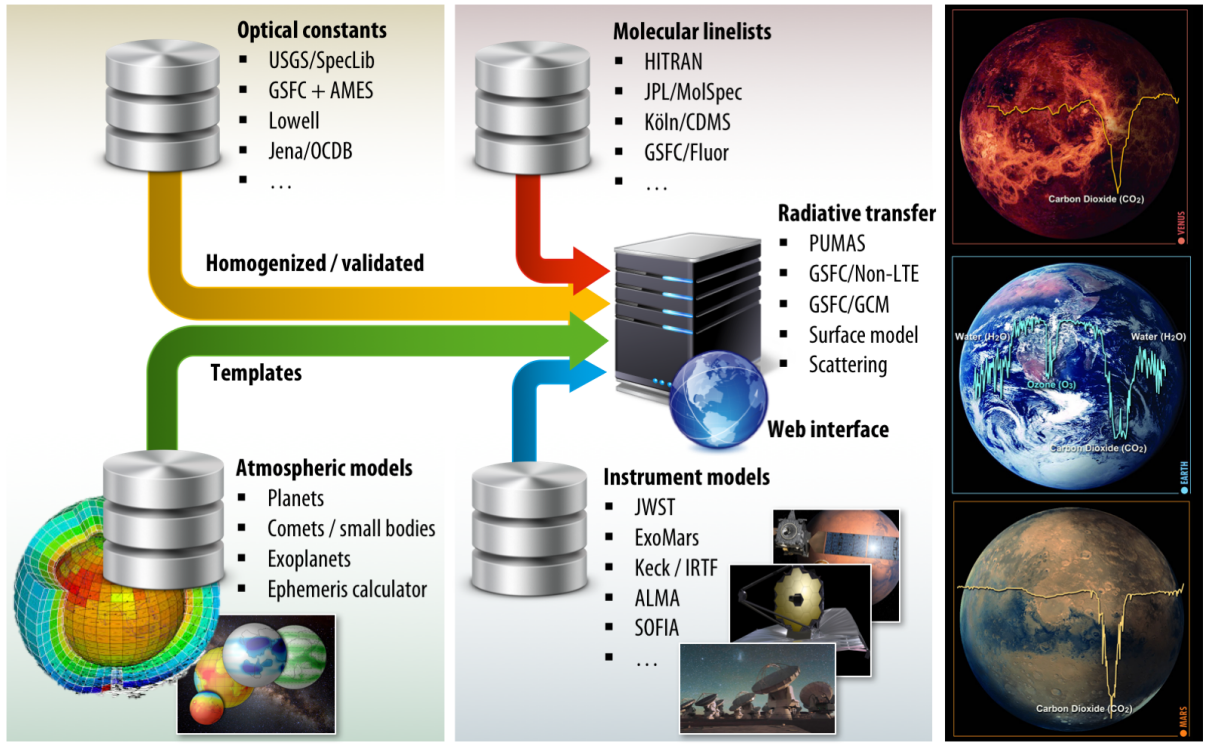

**Figure 1:** The Planetary Spectrum Generator (PSG) integrates a broad range of spectroscopic, atmospheric and instrument databases which are ingested by a versatile radiative transfer suite. By providing a user-friendly web interface to these models and databases, the tool can be utilized to synthesize broad range of planetary spectra. Right panel spectra based on [20] and ESA/Medialab.

As we discuss in chapter 2, the first step for any simulation is a precise three-dimensional description of the object under study and the geometry of the observation. In chapters 3 and 4, we present the models and databases employed to describe the atmospheres and surfaces of these bodies, respectively. In chapter 5, we discuss the different methods employed in performing the radiative-transfer models. A description of the online tool and the remote application program interface (API) is presented in chapter 6, while conclusions summarizing the tool and the future steps are presented in chapters 7 and 8.

## 2.0    Planetary bodies and geometries

Any spectroscopic simulation requires a precise description of the geometry being considered. For that purpose, we have integrated a three-dimensional (3D) calculator for most bodies in the solar system, and all confirmed exoplanets. By setting the appropriate observational geometry, PSG can synthesize spectra for practically any observer and target configuration (specific details on the observing parameters accepted by PSG can be found in Appendix A). These can be summarized as adaptions of observatory, nadir, limb, solar/stellar occultation or from-surface (looking-up or looking-up to the Sun) remote sensing observations (see figure 2). Once the 3D location of the object and the observer are defined, the geometry module computes the following set of parameters for the field-of-view (assumed to be a circular beam): a) incidence angle ($\beta$); b) emission angle ($\alpha$); c) phase angle between the observer and the Sun ($\phi$); d) fraction of the object included in the FOV (field-of-view); e) fraction of the parent-star included in the FOV (SP); f) projected distance between the FOV and the parent-star (DS). Certain parameters are only relevant to specific geometries, for example the parameters SP and DS are only relevant when observing exoplanets, and DS specifically when employing a coronagraph. In limb and occultation geometries, the incidence and emission angles are equal to 90 ($\beta=\alpha=90°$). In the looking-up mode, the radiative transfer is integrated only along the emission path ($\alpha$), while the $\beta$ and $\phi$ parameters are only used to compute the diffuse scattering fluxes.

When the FOV is much smaller than the object disk (e.g., nadir, limb, occultation and looking-up observations), a single set of geometry parameters is typically sufficient when performing the radiative transfer calculation. The issue is when the FOV samples a broad range of illuminations and surface properties (e.g., the FOV is comparable and/or bigger than the object disk); in this case, one would need to compute radiative transfer simulations over different geometries, which would be then integrated to produce a single total planetary flux. Such approach would be particularly computationally expensive, and in PSG we currently define only one set of geometry parameters and one radiative-transfer calculation per simulation (the user can decide to perform detailed mapped simulations by employing the application program interface [API], see section 6.1). In many cases, by establishing a single set of representative geometry parameters, one can obtain accurate total planetary fluxes from a single radiative transfer calculation, even when the FOV encompasses the whole planetary disk. In PSG, we employ a hybrid approach, in which the geometry module computes fluxes and geometry parameters (e.g., emission:$\alpha_i$ and incidence:$\beta_i$ angles) across the sampled FOV/disk employing a grid of 140 x 140 points (19,600 sets of geometry values). The point-by-point fluxes are computed employing a Lambertian model, and are then used to determine the contribution function ($w_i$) for each of these points to the total flux.

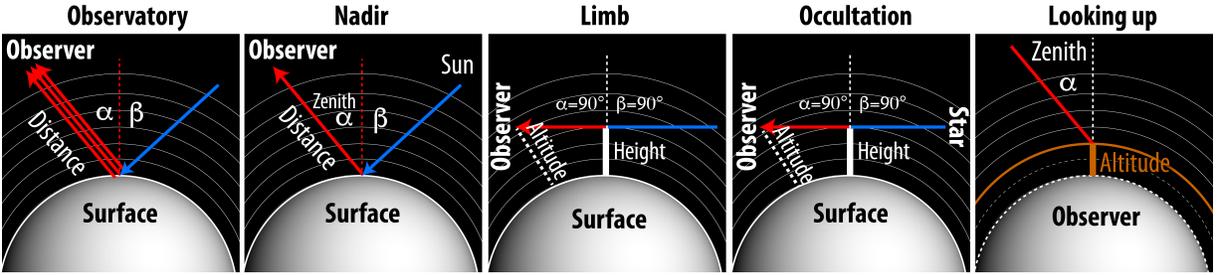

**Figure 2:** Types of geometries considered by PSG. Two main geometry parameters define the viewing configuration for each geometry: incidence angle ($\beta$) and emission ($\alpha$) angles for the observatory and nadir geometries; orbiter altitude and tangent height (i.e., impact parameter) for the limb and occultation geometries; observer's altitude (with respect to the planetary surface) and zenith angle for the "looking-up" geometry.

The effective emission angle ($\alpha$) and incidence angle ($\beta$) are then calculated as the flux weighted value across the FOV, $\alpha = \cos^{-1}(\Sigma(w_i \cdot \cos(\alpha_i)))$ and $\beta = \cos^{-1}(\Sigma(w_i \cdot \cos(\beta_i)))$. The same procedure is used to compute the effective sub-solar and sub-observer latitudes and longitudes.

## 2.1 Orbital modeling of solar system bodies

For the main bodies in the solar system (e.g., Mars, Neptune, Europa), PSG relies on pre-computed ephemeris tables from 1960 to 2050 calculated with the JPL/Horizons ephemerides system (https://ssd.jpl.nasa.gov/horizons.cgi) at a cadence of 1 hour, and later interpolated to the observational time, establishing the geometrical location and orientation of the object with respect to the observer. These tables tabulate heliocentric distance (rh), heliocentric velocity (vh), geocentric distance (rg), geocentric velocity (vg), sub-solar latitude (slat), sub-solar longitude (slon), sub-observer latitude (olat) and sub-observer longitude (olon). For small bodies (e.g., comets, asteroids), PSG dynamically extracts orbital parameters from the JPL-Horizons ephemerides system by connecting via their telnet API. For small bodies with no sub-solar / sub-observer latitudes and longitudes, PSG defines them based on the ecliptic angles.

A key climatological parameter is the solar longitude (Ls, see figure 3), or position across the orbital path (true anomaly) with respect to the northern hemisphere spring equinox (time when the planet's equatorial plane is equal to the orbital plane, Ls=0). Ls=90 corresponds to northern summer solstice, Ls=180 marks the northern autumn equinox and Ls=270 indicates northern winter solstice. This parameter is particularly relevant when querying information from general-circulation-models (GCM) or climatological databases of planets, since it is a fundamental factor in describing the structure and dynamics of the atmosphere. From the previously described ephemerides tables, which also include the true anomaly quantity, we determined the Ls=0 for each object and used this information to establish the seasonal state and climatology of the planet. When defining the reference point (i.e., location of the observer), PSG typically assumes an Earth-centered position, yet detailed spacecraft information (with a cadence of 1 minute, and frequently updated) has been integrated into PSG for many orbiters (e.g., MRO, Mars Express, ExoMars/Trace Gas Orbiter, MAVEN, Mars Odyssey, Cassini, Juno) and hundreds of points of interest (e.g., Maunakea, Paranal, Chajnantor, Arecibo, Viking, MSL/Curiosity, MER landers).

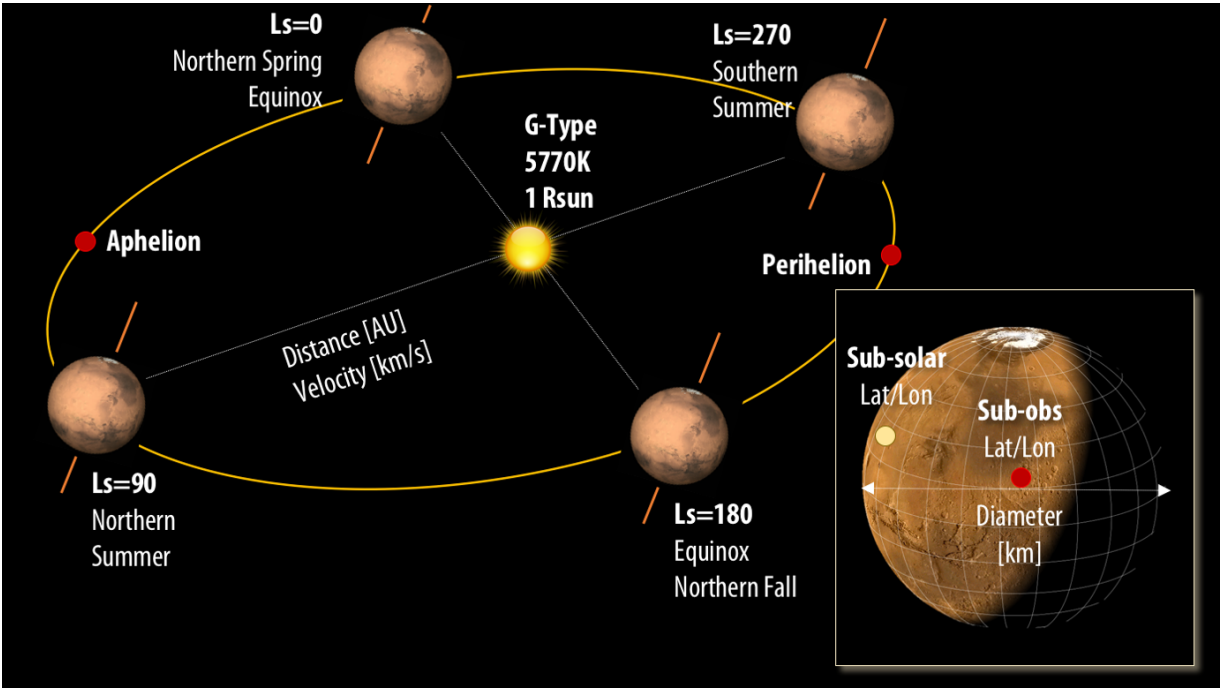

**Figure 3:** Description of the main orbital parameters used by PSG to determine the three-dimensional geometry of the observation – Mars shown in this example. The sub-observer latitude and longitude defines the center location of the planetary disk as viewed from the observer, while the sub-solar latitude and longitude defines the center as viewed from the parent star. The solar longitude (LS) is a proxy to the seasons on the planet and defines the position across the orbital path (true anomaly) with respect to the northern hemisphere spring equinox (time when the planet's equatorial plane is equal to the orbital plane, Ls=0).

## 2.2 Orbital modeling of exoplanets

Transit detection and radial velocity characterization of exoplanet orbits provide constraints on the orbital parameters of planets detected around other stars. This information can be used to construct a three-dimensional view of the system and is used by PSG to predict the time of future primary and secondary transits. The NASA Exoplanet Archive (https://exoplanetarchive.ipac.caltech.edu) is an excellent repository of the constantly growing database of exoplanet orbital parameters. For exoplanets, PSG does not employ ephemerides tables, but computes the orbital integration of the Keplerian orbit using parameters obtained from the NASA Exoplanet Archive (via their API). This ensures that PSG uses the latest and most accurate orbital parameters, and that newly discovered exoplanets can be properly handled by PSG. Unfortunately, Kepler's equation that relates the true anomaly (T, position on the elliptical orbital path) and the mean anomaly (M, proportional to time) does not have a closed-form solution (i.e., transcendental equation), and the orbital calculation requires an iterative solver. PSG employs the iterative Newton–Raphson algorithm to determine the true anomaly (T), constrained to an eccentric anomaly (E) precision of 1E-6 radians and to a maximum of 30 iterations. This computation employs the longitude of periapse (w), orbital eccentricity (e) and time of transit (TT) or time of periastron (TP). Even for highly eccentric orbits, such a method is sufficient to accurately determine the location of the exoplanet along its orbit, and this together with knowledge of the orbital inclination (I), permits PSG to construct an accurate three-dimensional view of the system (within the known certainty of these variables).

Determining the actual sub-solar and sub-observer latitudes / longitudes requires knowledge of the rotational period and obliquity of the planet. Such parameters are rarely known for exoplanets, and when computing exoplanet ephemerides PSG assumes that the planets are tidally locked with no obliquity and the sub-solar latitudes / longitudes are set at the center of the planet, while the phase identifies the true anomaly with respect to that of the secondary transit (with a phase of 180 corresponding to the primary transit). Arbitrary or user-defined sub-solar and sub-observer values can be edited manually in the configuration file, or via an API request (see Section 6.1).

### 3.0 Atmospheres and line-lists

The main structure of an atmosphere can be assumed to be in hydrostatic equilibrium (in which atmospheric pressure is equilibrated by gravity), or in constant expansion (typical for comets and small bodies), in which gravity is negligible (see figure 4). When establishing the vertical structure of the atmosphere (T vs. pressure, abundance profiles, etc.) the user can provide any arbitrary vertical profile as input by modifying the "ATMOSPHERE" fields (see Appendix A) of the configuration file. As a minimum, the overall structure of the atmosphere can be defined with a set of three parameters: temperature ($T_{atm}$ [K]), surface pressure ($P_{surf}$ [bar]) and molecular weight ($m_{atm}$ [g/mol]) for equilibrated atmospheres or production rate ($Q_{coma}$ [s⁻¹]), temperature [K] and expansion velocity ($v_{exp}$ [m/s]) for expanding ones. In such a simplified case for equilibrated atmospheres, the temperature (T) is assumed to be constant across altitude, while the pressure (P [bar]) decreases with altitude (z [m]) following the scale-height (H [m]): $P = P_{surf} \exp(-z/H)$, where $H=RT_{atm}/(m_{atm}g)$, g is the surface gravity (m/s²), R is the gas constant ($8.3144598E3$ g m² s⁻² K⁻¹ mol⁻¹). Further parameterization is permitted in PSG, and the radiative transfer also allows the user to provide layer-by-layer pressures, temperatures, altitudes and abundances. For most of planets in the solar system, PSG provides vertical profile information based on climatological databases (see sections 3.1 and 3.2), equilibrium models (see sections 3.5), and remote-sensing and in-situ measurements (e.g., Venus [21], Jupiter/Saturn [22], Neptune/Uranus [23] and Titan [24]). Computation of layer-by-later integrated column densities (molecules/m²) and aerosols mass densities (kg/m²) is done employing the Curtis-Godson algorithm [25,26].

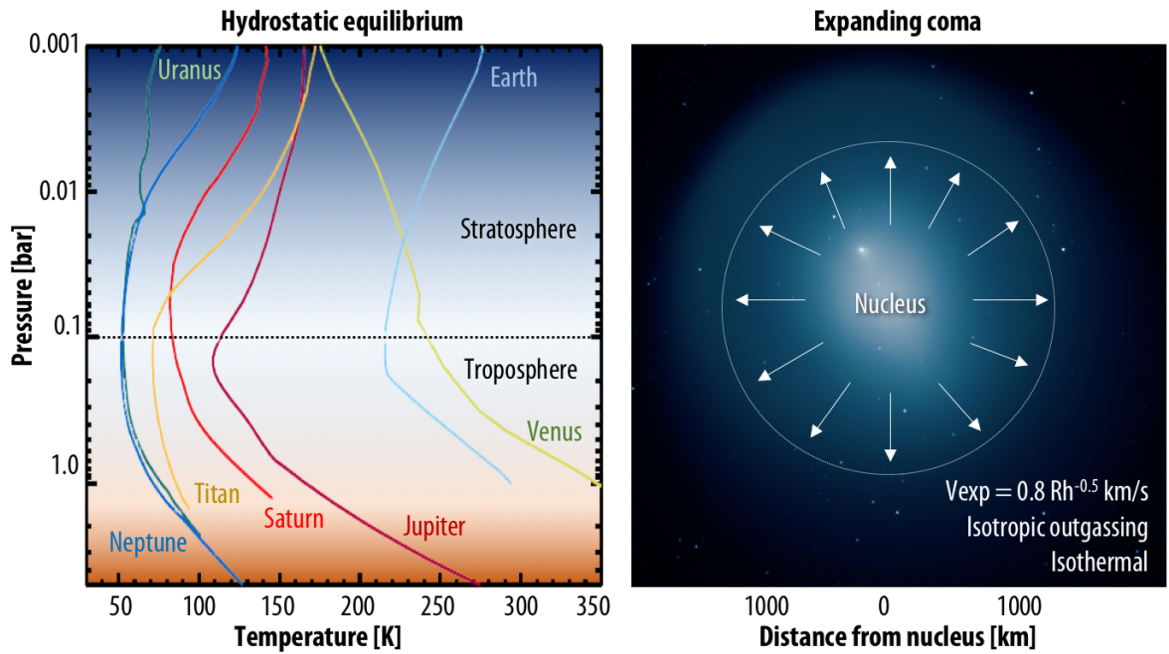

**Figure 4:** PSG operates with two basic types of atmospheres: hydrostatic equilibrium (typical for planets, profiles from [27]) and expanding coma (typical of comets and small bodies). For the main planets, vertical profiles and climatological information is available, while the user can also load any arbitrary vertical structure. For expanding atmospheres, PSG assumes isotropic outgassing, a constant temperature across the coma, and an outgassing velocity established by the heliocentric distance.

Even though the user may provide vertical profile information for a broad range of species and aerosols, in many cases, only a selected set of species are needed to be included in the radiative-transfer analysis. More importantly, it is of key importance that the user defines which molecular database will be used to synthesize each component, and for that purpose PSG permits the user to link the species to be analyzed with the specific spectral database. There are several highly complete spectral databases, each applicable to different spectral regions and excitation regimes (see summary table 2). For instance, HITRAN [14] and GEISA [15] are mostly complete (in the IR, optical and UV at moderate temperatures) for most typical planetary atmosphere and have become the main repositories of line information, while at radio wavelengths the JPL Molecular Spectroscopy [16] and Cologne Database of Molecular Spectroscopy (CDMS, [17]) are generally more complete and have a better description of the rotational spectrum of complex molecules. NASA-Goddard currently holds the main repository for non-LTE fluorescence linelists, suitable when synthesizing cometary spectra in the UV/optical/IR range [28–32]. The JPL and CDMS databases do not provide lineshape information, and when using spectroscopic information from these to synthesize planetary spectra, PSG assumes standard lineshape parameters (Lorentz [HWHM]=0.07 $cm^{-1}$, a temperature-dependence exponent for the line-width of 0.75, and no pressure shift, see [14] for details). For aerosols, PSG allows the user to select from a broad selection (103 species) of pre-computed scattering models (see section 5.1), that range from icy particles to sulfuric volcanic aerosols. These models are tabulated at four particle sizes (0.01, 0.1, 1 and 10 $\mu m$), and later interpolated to the desired user provided particle size.

| Table 1. Database of linelists handled by PSG | Spectroscopic capabilities |
|---|---|
| HITRAN 2016 [14] | **Wavelength range**: 0.3 $\mu m$ to radio<br>**Number of lines**: 5,399,562<br>**Number of molecules**: 50<br>**Number of isotopologues**: 126<br>**CO$_2$/H$_2$/He broadening**: 13 molecules<br>**Number of CIA spectra**: 554<br>**Number of cross-section spectra:** 987<br>**Number of aerosols**: 98 |
| GSFC Fluorescence Database [28–32]<br>(C$_2$H$_2$, C$_2$H$_4$, C$_2$H$_6$, CH$_3$D, CH$_3$OH, CH$_4$, CO, CO$_2$, H$_2$CO, H$_2$O, H$_2$S, HC$_3$N, HCN, HDO, HNC, NH$_3$, OCS) | **Wavelength range**: shorter than 10 $\mu m$<br>**Number of lines**: 530,281<br>**Based on lines** (non-LTE): 2 billions<br>**Number of species**: 17 |
| GEISA 2015 [15] | **Wavelength range:** 0.3 $\mu m$ to radio<br>**Number of lines**: 5,023,277 |
| | **Number of species:** 52<br>**Number of isotopologues:** 118 |
| JPL Molecular Spectroscopy [16] | **Wavelength range:** 2.65 µm to radio<br>**Number of lines:** 888,113<br>**Number of species**: 383 |
| CDMS Cologne Database for Molecular Spectroscopy [17] | **Wavelength range:** 1.81 µm to radio<br>**Number of lines:** 1,612,154<br>**Number of species:** 792 |
| Exo-Transmit Opacities database [33] | **Wavelength range:** 0.1 to 170 µm<br>**Number of spectral points:** 7454<br>**Number of temperatures:** 30 (100 to 3000 K)<br>**Number of pressures:** 13 (1E-9 to 1000 bar)<br>**Number of species:** 30 |
| The MPI-Mainz UV/VIS Spectral Atlas [34]<br>UV cross-sections for O3 [35]<br>UV cross-sections for CO2 [36] | **Wavelength range:** 0.01 to 1 um<br>**Number of cross-sections:** 70<br>**Number of species:** 22 (e.g., $H_2O$, $CH_4$, $CO_2$, $N_2O$, $O_3$) |

## 3.1 Earth atmosphere

For the computation of telluric transmittances and radiances, PSG extracts atmospheric profiles and molecular abundances from the Modern-Era Retrospective Analysis for Research and Applications (MERRA-2) database [37]. MERRA-2 is the latest atmospheric reanalysis of the modern satellite era produced by NASA's Global Modeling and Assimilation Office, which incorporates information from hundreds of orbiters and ground stations since 1980 and provides global three-dimensional of atmospheric parameters (e.g., temperature, abundance profiles, aerosols). Specifically, PSG works with the M2I3NVASM component, which provides assimilated meteorological fields (pressure, temperature, water vapor, ozone, and water ice clouds) from the surface to ~80 km (72 layers) with a cadence of 180 minutes, and spatial resolution of ~0.5 degrees (576 x 361). The values are further refined temporally and spatially to a resolution of better than 1 km employing the USGS-GTOPO30 topographic maps [38] and considering a hydrostatic equilibrated atmosphere within every bin. The MERRA-2 database is constantly being updated, yet it assimilates data with a latency of approximately 4 months. For atmospheric requests before 1980, or newer than present minus 4 months, PSG will extract information for the same season / location / time-of-day but for the nearest year available in the database.

For species not contained in this repository ($CO_2$, $N_2O$, $CO$, $CH_4$, $O_2$, $NO$, $SO_2$, $NO_2$, $NH_3$, $HNO_3$, $OH$, $HF$, $HCl$, $HBr$, $HI$, $ClO$, $OCS$, $H_2CO$, $HOCl$, $N_2$, $HCN$, $CH_3Cl$, $H_2O_2$, $C_2H_2$, $C_2H_6$, $PH_3$), PSG will consider standard abundance profiles [13]. Many of these species have long atmospheric lifetimes (e.g., $CO_2$, $CH_4$) and are non-condensable (e.g., $CO$) so they do not show strong geographical and/or temporal variability, so this assumption should be accurate enough in most cases. Yet other species are known to display strong variability (e.g., $H_2CO$) and the provided vertical profiles should be considered solely as best-guesses.

## 3.2 Mars atmosphere

PSG extracts atmospheric information from the Mars Climate Database (MCD, [19]), a comprehensive repository of meteorological fields derived from General Circulation Model (GCM) numerical simulations and validated using available observational data. The database has a natural spatial resolution of ~5 degrees (64 x 49) and 30 layers up to ~90 km, with 12 local times (2-hours resolution) and 12 seasons. PSG operates with the raw information contained in the NetCDF files, which is later refined spatially and temporally to the specific time and latitude / longitude of the PSG calculation. In a similar procedure as done by Millour et al. [19] (and as we do to resample the MERRA-2 database), the 64 x 49 grid is refined spatially to a much finer scale (11520 x 5760, ~0.03 degrees) by employing the MOLA topographic data [39]. Several climatological states are possible (e.g., dust storm, cold scenario) and also for different levels of EUV radiation, with PSG extracting information only considering the typical climatological state (average dust and EUV conditions). Specifically, PSG extracts vertical profiles of pressure, temperature, $CO_2$, $N_2$, $O_2$, $CO$, $H_2O$, dust, dust size, water ice and ice particle size from the MCD database, while the surface albedo information is determined from the MGS/TES database [40].

## 3.3 Gas giants / Dense and hazy atmospheres

Planets in which their atmospheres are so thick, that the definition of a physical solid "surface" becomes ambiguous (e.g., Jupiter, Saturn, Uranus, Neptune), require a careful and particular parameterization of the radiative-transfer. Specifically, the atmospheric layering should be as complete as possible when defining the main species that contribute to the spectroscopic opacities. In giant planets, the main opacity terms that define this optically thick "surface" are hazes (e.g., $CH_4$, $NH_3$, $NH_4SH$, $H_2S$, $H_2O$ hazes) and the effect of collision-induced-absorption by molecular hydrogen and helium. Beyond that, a plethora of hydrocarbon species (e.g., $CH_4$, $C_2H_2$, $C_2H_6$) and other trace molecules contribute to the overall atmospheric opacities. Due to this set of numerous opacity terms, these atmospheres are rarely probed at pressures beyond 10 bars, and in PSG we have provided a basic description of the main hazes and species from 10 bars to ~10 nbar. For the giants (i.e., Jupiter, Saturn, Uranus and Neptune), the haze profiles are based on [41] while for Venus, the sulphuric acid haze profile is based on [42]. Molecular abundances are only prescribed for a select set of species based on [22,23,43,44], and it is great of importance for the user to update and validate the desired abundance / haze profiles when performing a simulation. Information about molecular lineshapes for non-terrestrial environments (e.g., $H_2$, He, $CO_2$ atmospheres) is still limited, yet with the advent of new laboratory measurements [45,46] and new modeling efforts [28,47], many more linelists are constantly added. Currently, PSG handles broadening coefficients for self-broadening and for an air (mixture of $N_2+O_2$) atmosphere for all HITRAN species, and for a selection of species ($H_2O$, CO, $SO_2$, HF, HCl, OCS, $C_2H_2$, $CO_2$, $N_2O$, $H_2CO$, HCN, $H_2S$, OH) in $H_2$, He and $CO_2$ atmospheres. We expect to provide broadening information for many more species in the near future (see section 7).

### 3.4 Cometary atmospheres

Cometary outgassing and molecular abundances are highly variable and relatively difficult to estimate or predict. The main parameter that establishes the intensity of spectroscopic fluxes in a comet is the gas production rate, or activity rate, Q [molecules/s]. The user can provide this value as a parameter, but in some cases the user may not know it, and PSG will estimate the level of activity in the comet (i.e., Q) from other observed astronomical quantities (e.g., visual magnitude). Typically, cometary activity is driven by water outgassing at heliocentric distances (rh) within 2 au, while the sublimation of more volatile ices tends to dominate cometary activity beyond 2 au, yet this is extremely dependent on composition and structure. Within 2 au, activity (Q [molecules/s]) is dominated by insolation, with outgassing rates following a $rh^2$ relationship (Q = $Q_{au}/rh^2$, where $Q_{au}$ is the activity at rh=1 au). Outgassing velocities also follow a heliocentric distance dependence, yet much less steep, and in PSG we employ an empirically defined relationship (valid in the collisionless region of the coma, beyond ~100 km from the nucleus) to establish the expansion velocities: $v_{exp}$=0.8·$rh^{-0.5}$ [km/s] [48,49]. In a related fashion and employing a sample of 37 comets observed from 1982 to 2004 (234 points), Jorda et al. [50] determined an empirical relationship between visual magnitude ($m_v$) and gas production rate (Q), $m_v$ = (30.675 - log(Q))/0.2453 + 5·log(rg). Cometary visual brightness is mainly defined by dust brightness, and such a relationship intrinsically implies a common gas to dust ratio among comets. Even considering this caveat, such a relationship is of great value when estimating cometary activity, and in PSG we use this to estimate the water production rates from visual magnitudes derived from JPL Horizon (see 2.1), yet the user can edit this estimate as needed.

Specifically, PSG computes three brightness metrics for cometary coma, which are displayed next to the production rate field: the expected visual magnitude ($m_v$), the infrared Figure-Of-Merit ($FOM_{IR}$), and the radio Figure-Of-Merit ($FOM_{Radio}$). These parameters are calculated based on the heliocentric distance (rh [au]), the geocentric distance (rg [au]), and the gas production rate (Q [molecules/s]), employing these equations: $m_v$ = (30.675 - log(Q))/0.2453 + 5·log(rg) [50]; $FOM_{IR}$ = Q·1E-29/(rg·rh$^2$) [51]; $FOM_{Radio}$ = Q·1E-28/rg [52]. Molecular abundances relative to Q tend to vary substantially between comets [32,53,54], yet a "typical" set of abundance ratios can be assumed, and in PSG we consider as a-priori: $H_2O$/ $CO_2$/ CO/ $CH_4$/ $NH_3$/ $H_2CO$/ $C_2H_6$/ HCN/ $CH_3OH$ = 100/ 15/ 12/ 1.5/ 0.5/ 0.3/ 0.6/ 0.4/ 2.4.

## 3.5 Exoplanet atmospheres

The temperature, composition and structure of exoplanets vary substantially, from scorching hot giants to small rocky planets. In an attempt to provide a moderately accurate description of this diversity, in PSG we organize the exoplanets by mass (M in Earth's mass [M⊕]) and density ($\rho$ [g/cm$^3$]) into four categories (in a comparable fashion to the Kepler categorization): Earth-like planets (M<2 and $\rho$>4), super-Earths (M<10 and $\rho$>4, bigger than Earth but smaller than Neptune), Neptune-like (M<50, small gas giants, comparable to Neptune and Uranus), and gas-giants (M>50, Saturn-sized and larger). We are currently working with exoplanet atmospheric modelers to develop a flexible module for describing Earth-like and super-Earth atmospheric profiles. Currently PSG simply assumes a terrestrial structure and composition (MERRA-2, see 3.1) for Earth-like planets, and a Venusian structure and composition [21] for super-Earth like planets. For gas giants, PSG employs the non-grey thermal model by Parmentier & Guillot [55] to determine the vertical temperature profile of the planet. This analytical model is fast and matches full numerical simulations within 10% over a wide range of effective temperature, internal temperature and gravity and properly predicts the depth of the radiative/convective boundary. The equilibrium temperatures of the atmospheres are assumed to be strongly related to their composition and to the acting greenhouse gases. The model makes reasonable assumptions about composition based on gravity, distance to host star and density, and using two different opacity bands in the thermal frequency range, establishes the dual role of thermal non-grey opacities in defining the temperature profile. Opacities dominated by lines enable the upper atmosphere to cool down significantly compared to a grey atmosphere whereas opacities dominated by bands lead both to a significant cooling of the upper atmosphere and a significant heating of the deep atmosphere [55]. Once the temperature profile of the exoplanet is defined, PSG employs the chemical equilibrium equations-of-state (EOS) derived by Kempton et al. [33] for a wide range of metallicities and primordial states. This implementation provides a-priori vertical profiles for dozens of species: $H_2O$, $CO_2$, $O_3$, $N_2O$, $CO$, $CH_4$, $O_2$, $NO$, $SO_2$, $NO_2$, $NH_3$, $HNO_3$, $OH$, $HF$, $HCl$, $HBr$, $HI$, $ClO$, $OCS$, $H_2CO$, $HOCl$, $N_2$, $HCN$, $CH_3Cl$, $H_2O_2$, $C_2H_2$, $C_2H_6$ and $PH_3$.

## 4.0 Modeling of planetary surfaces

The surface defines one of the "boundary" conditions of the radiative-transfer analysis. For expanding atmospheres, these parameters relate to those of the nucleus and to the dust particles (solid phase components), with the parameter "dust/gas" ratio identifying whether the comet is dust poor (low dust/gas) or rich (high dust/gas). We have developed a versatile surface module that combines a realistic Hapke scattering model [6] and the capability to ingest a broad range of optical constants, permitting PSG to accurately compute surface reflectances and emissitivities. This type of modeling is applicable to surfaces both in the inner and the outer solar system and it is of high relevance to studies related to formation and evolutionary processes over a wide range of distances from the Sun (e.g., [56–59]).

When synthesizing planetary reflectance spectra, a comprehensive and validated repository of spectral constants and reflectance spectra is mandatory in order to perform this task. The spectroscopic signatures are affected by the relative abundance of the components together with their grain sizes, and the approximate surface characteristics (e.g. mean roughness slope, compaction parameter) of the layer that scatters the incident solar radiation. Real planetary surfaces are composed of different elements, and several methods have been developed to calculate synthetic reflectance spectra for comparison with the spectroscopic observational data of solar system bodies. One of the scattering theories most widely used for modeling the reflectance spectra of planetary surfaces is that of Hapke [60,61]. This geometric optics model provides the bidirectional reflectance $r$ of a particulate surface as a function of the single scattering albedo $\omega$. The latter is computed by PSG using the equivalent-slab approximation and the bidirectional reflectance is described as $r = (\omega(D,n,k)/4\pi) \, (\mu/(\mu + \mu_0)) \, P(g)$. When employing optical constants ($n$ and $k$), the volume single scattering albedo $\omega$ is calculated as a function of the particle diameter $D$, and the formula above is applied. However, the user can also select reflectance measurements to represent the surface, instead of optical constants, in which case the formula above is not needed. The cosine of the incidence angle ($i$) and emission angle ($e$) are commonly indicated with $\mu_0$ and $\mu$, respectively. The particle phase function $P(g)$, with $g$ being the phase angle, accounts for anisotropic scattering. A one-term Henyey-Greenstein phase function is considered for $P(g)$.

The previous relation can be used to compute the bidirectional reflectance of a medium composed of closely packed particles of a single component. However, the surface of interest can be a mixture of different types of elements. Therefore, in order to calculate synthetic reflectance spectra for comparison with the observational data, it is necessary to compute the reflectance of mixtures of different types of particles. An areal (also called geographical) mixture consists of materials of different composition and/or microphysical properties that are spatially isolated from one another, while in an intimate mixture the surface consists of different types of particles mixed homogeneously together in close proximity. Areal is the most commonly considered approach and it is also substantially less computationally intensive; therefore it is the current method employed by PSG.

Properly modeling spectroscopic features of planetary surfaces over a wide wavelength range requires a comprehensive and inclusive spectroscopic database. There is currently no single repository that integrates optical constants and reflectances of solid surfaces and ices over a wide spectral range. For instance, for interpreting data by CRISM (0.4-4 µm, Compact Reconnaissance Imaging Spectrometer for Mars instrument on MRO), the team has created a repository of spectral constants (see MRO/CRISM listed below), while the New-Horizons teams utilize several specialized optical constants. Each database has a unique file system and nomenclature, making it extremely difficult to integrate different spectral constants into a common surface radiative transfer model.

We identified eight spectral databases (see Table 4) containing information for 1974 materials that are applicable to the synthesis of planetary spectra, and integrated them into PSG. In order for a single radiative transfer tool to be able to integrate information from these databases, a standardized and homogenous library of spectral constants was created. The process of validation and homogenization involves removing invalid entries (e.g., negative reflectances, invalid numbers), deleting duplicate entries, and most importantly converting the values into standard and portable ASCII column tables (wavelength vs. reflectance/constants) across all repositories. The next step was compiling an easy to search summary table with enough information about each measurement (e.g., wavelength range, temperature, material) so the radiative transfer can properly incorporate these parameters into the planetary calculation.

---

**Table 2. Database of surface repositories handled by PSG**

Mars Reconnaissance Orbiter (MRO) Compact Reconnaissance Imaging Spectrometer for Mars instrument (CRISM) spectral library: repository of spectral templates applicable to Mars [62]
**Number of components**: 31
**Spectral coverage**: 0.44 to 3.9 μm
**Type of parameters:** reflectances (PSG type 0)
**Online repository**: http://crismtypespectra.rsl.wustl.edu

---

United States Geological Survey (USGS) digital spectral library (version splib06a): repository of a wide range of materials and components [63]
**Number of components**: 1380
**Spectral coverage**: most in the 0.3 to 3 μm range, with some reaching ~200 μm.
**Type of parameters:** reflectances (PSG type 0)
**Online repository**: http://speclab.cr.usgs.gov/spectral.lib06/ds231/datatable.html

---

Database of Optical Constants for Cosmic Dust (DOCCD): repository of optical constants for a wide range of silicates, oxides, sulfides, carbonates, carbides and carbon materials [64]
**Number of components**: 106
**Spectral coverage**: most in the 2 to 500 μm range, with some reaching ~10000 μm.
**Type of parameters:** optical constants (PSG type 1)
**Online repository**: http://www.astro.uni-jena.de/Laboratory/OCDB

---

Lowell Observatory Grundy's database of optical data on cryogenic ices ($N_2$, $H_2O$, $CH_4$) [5]
**Number of components**: 3
**Spectral coverage**: 0.6 to 5 μm
**Type of parameters:** alpha parameter (PSG type 2)
**Online repository**: http://www2.lowell.edu/users/grundy/ice.html

---

NASA Goddard's Cosmic Ice Laboratory: optical constants for selected ices of astrobiological relevance [65]
**Number of components**: 22
**Spectral coverage**: 2 to 20 μm and 2 to 333 μm
**Type of parameters:** optical constants (PSG type 1)
**Online repository**: http://science.gsfc.nasa.gov/691/cosmicice

NASA Ames' Database of Astrochemical Ices [66]
**Number of components**: 6
**Spectral coverage**: 2 to 200 μm, 2 to 20 μm, 3 to 14 μm
**Type of parameters**: optical constants (PSG type 1)
**Online repository**: http://www.astrochem.org/db.php

HITRAN Refractive index repository [67]
**Number of components**: 97
**Spectral coverage**: wide range from UV to radio
**Type of parameters**: optical constants (PSG type 1)
**Online repository**: http://hitran.org

Bus-DeMeo asteroid taxonomy with 25 classes based on PCA of combined visible and near-IR spectral data [68]
**Number of components**: 25
**Spectral coverage**: 0.45 to 2.45 μm
**Type of parameters:** reflectance spectra (PSG type 0)
**Online repository**: http://smass.mit.edu/busdemeoclass.html

## 5.0 Radiative transfer modeling

The spectroscopic calculation is divided in three stages within PSG: 1) calculation of the surface reflectance / emissivity and solar spectra, 2) calculation of the atmospheric radiances and transmittances, 3) calculation of the observable fluxes as convolved with the instrument / telescope transfer function. For the first stage, we have developed an efficient surface scattering model that ingests a wide range of reflectances and optical constants (see section 4), while for the second stage PSG employs two radiative-transfer models (PUMAS and CEM, see figure 5). PUMAS, or Planetary and Universal Model of Atmospheric Scattering [1], is the layer-by-layer radiative transfer model employed by PSG for computing spectra of hydrostatic equilibrium atmospheres. It integrates the latest radiative-transfer methods and spectroscopic parameterizations, in order to compute high resolution spectra via line-by-line calculations, and utilizes the efficient correlated-k method at moderate resolutions. The scattering analysis is based on a Martian scattering model [69], while the line-by-line calculations have been validated and benchmarked with the accurate GENLN2 model [12]. For computing cometary spectra, PSG employs CEM, or Cometary Emission Model [28,30,70], which incorporates excitation processes via non-LTE line-by-line fluorescence model (employing GSFC databases), and ingests HITRAN, GEISA, JPL and CDMS spectral databases to compute line-by-line LTE fluxes. It operates with expanding coma atmospheres, and computes photo-dissociation processes for parent and daughter species released in the coma.

Importantly, PSG includes the possibility of integrating stellar spectroscopic templates by adopting the Kurucz stellar templates (0.15-300 μm, [71]) and spaceborne high-resolution (0.02 cm$^{-1}$) solar data - the UV stellar signatures are highly variable and should be taken solely as a reference point. This stellar information is used to compute reflected/scattered solar/stellar fluxes, and also to compute the total observable exoplanet fluxes. The stellar transmittance templates can be also scaled to different effective stellar temperatures and when considering the G-type template the Kurucz spectrum is complemented by the ACE solar spectrum (2-14 μm, [72]). Each spectrum is properly shifted by the corresponding Doppler shifts (object-Star, object-observer), and PSG includes correction for rotational shift (spectral shift induced by the rotation of the planet) and rotational line broadening.

## 5.1 Radiative transfer of atmospheres (PUMAS)

When performing the line-by-line radiative-transfer analysis, PUMAS computes the layer-by-layer contribution of each line considering a Voigt line-shape implemented via a Faddeeva function with an optimized Humlicek algorithm [73,74]. This implementation, while fast, can be extremely computer intensive when computing the contribution from the extended Lorentzian wings. For that purpose, in PSG we employ the wide/fine grid methodology implemented in the LINEPAK [75] and GENLN2/3 [12], in which the core of the line is computed at a fine spectral resolution (dv=1E-3 cm$^{-1}$) while the wings are computed at a wider resolution (dv=1 cm$^{-1}$) with a typical wing extension of ±25 cm$^{-1}$.

When the spectral region of analysis includes millions of lines, a line-by-line analysis is prohibitively expensive and PUMAS employs the alternative correlated-k approach (e.g., [76]). This method is accurate for moderate resolutions, yet it requires to have pre-computed opacity tables for a broad range of species, temperatures, pressures, wavelengths and resolutions, and can be a daunting process to implement for such a generalized radiative-transfer suite as PUMAS. Currently, the correlated-k implementation in PUMAS has been only implemented for most of the HITRAN species (see table 2), for four collisional regimes (air, $H_2$, He and $CO_2$), for the complete PSG spectral range (0.1 μm to 100 mm) at two spectral resolving powers (5000 and 500), for 17 pressures (100 bars to 1 μbar) and 20 temperatures (40 to 2000 K), and 20 k-bins. Correlated-k bins are derived for each layer by interpolation, while the radiative transfer is always computed at the correlated-k resolution and further down sampled to the user's grid. When the requested spectral resolution is higher than 5000, PSG employs the above-mentioned line-by-line method. As an alternative to correlated-k, PUMAS can also ingest cross-section tables tabulated for different pressures, temperatures and collisional partner. Currently, PUMAS has access to hundreds of cross-sections for trace species as contained in the latest HITRAN release (see section 3), and for 30 species relevant to exoplanet research as compiled by Kempton et al. 2017 (see section 3). Collision induced absorptions (CIA) are also of key relevance in high pressure regimes, and PUMAS incorporates this phenomenon in the radiative transfer calculations by ingesting the 554 cross-sections reported in HITRAN (see section 3) for dozens of interactions (e.g., $O_2$-$O_2$, $O_2$-$CO_2$, $H_2$-He, He-H).

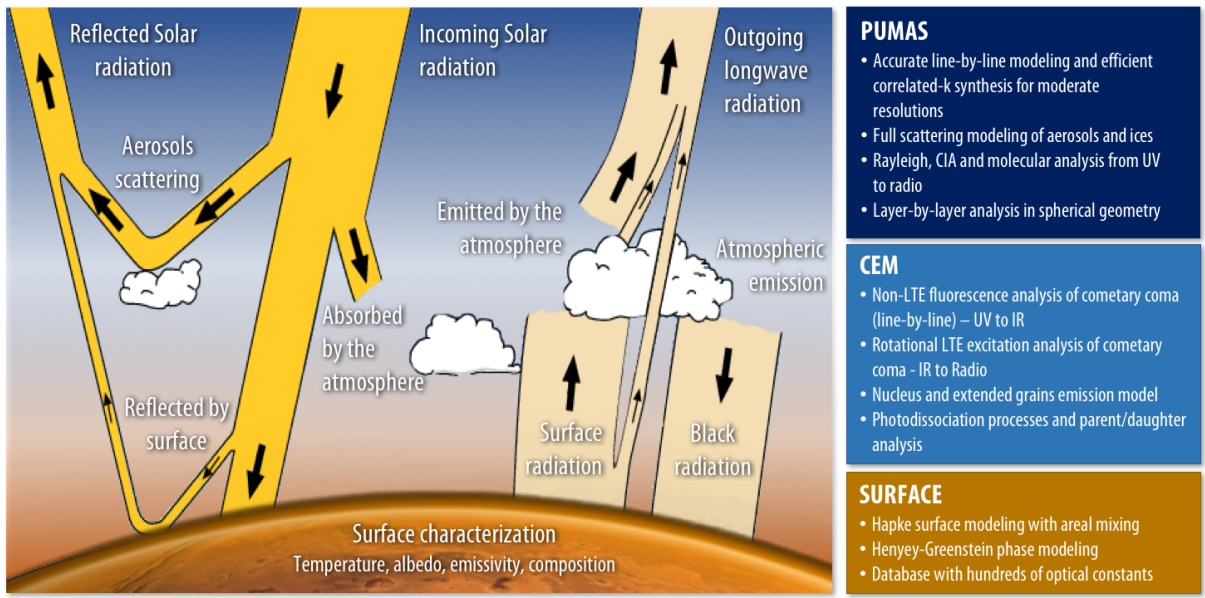

**Figure 5:** The diagram shows the different components considered by the radiative transfer modules. By performing a layer-by-layer analysis, PUMAS integrates and calculates the different flux contributions across the wavelength grid and atmosphere. For comets, the molecular calculation is performed separately by CEM from the surface fluxes, and later added to compute integrated fluxes.

Beyond molecules, the presence of aerosol particles (dust/ice/clouds/hazes) in a planetary atmosphere has a significant impact on the intensity and morphology of planetary spectra. The general approach of the community has involved the coupling of a molecular transmission code (i.e., LBLRTM, GENLN3) with a multiple scattering radiative transfer algorithm (i.e., DISORT, [77,78]) such as has been done by Turner and collaborators with LBLDIS [79]. In PSG, the treatment of multiple scattering from atmospheric aerosols is enabled by using the discrete ordinates method (e.g., [69,80,81]). The radiation field is approximated by a discrete number of streams distributed in angle with respect to the plane-parallel normal. The number of stream pairs (pairs of corresponding upward and downward radiation streams) can be set as high as necessary to accurately model the angular dependence of the aerosol scattering phase function while maintaining computational feasibility. The "two-stream approximation" often used when modeling planetary atmospheres is an example of the discrete ordinates method using one stream pair.

The angular dependence of the scattering phase function for a particular aerosol is described in terms of an expansion in terms of Legendre Polynomials, typically with the number of expansion terms equal to the number of stream pairs. As implemented in PSG, the Legendre expansion coefficients are pre-computed using an assumed particle size distribution for each available aerosol type and using either Mie scattering (e.g., [82]) or T-matrix (e.g., [83]) codes as specified in the associated information files for each aerosol type. The underlying indices of refraction for aerosols are empirically derived from spacecraft observations in the case of Mars dust and water ice aerosols [84], or from the HRI (HITRAN Refractory Index [67]) database. For the case of the HRI constants, we calculate scattering coefficients employing a Mie implementation [85] that derives Henyey-Greenstein scattering g-factors. Internally, PSG converts these g-factors into Legendre expansion coefficients in order to be ingested by the radiative-transfer suite.

The discrete ordinates formulation computes the diffuse radiation field for a plane-parallel atmosphere. When spherical geometry is important (e.g., limb geometry observations), the pseudo-spherical approximation (e.g., [81,86]) is used for computational efficiency. In this scheme, the source functions computed using the diffuse field from the discrete ordinates plane-parallel geometry are integrated along an equivalent curved path through the model layers. This curved path is defined by computing the correct emission angle for the path at the boundary of each layer. The pseudo-spherical approximation is accurate over a wide range of conditions and is orders of magnitude faster than an "exact" Monte Carlo code (see details in [69]). When computing transit spectra, the radiative transfer computes the slant transmission across every layer, and this is then integrated across all layers, ultimately deriving the effective transit exoplanet spectrum [21].

## 5.2 Cometary modeling (CEM)

The spectra of a comet can be divided into three main components: the nucleus (reflected and emitted), coma grains (reflected and emitted) and gas emissions (e.g., LTE, non-LTE). The Cometary Emission Model (CEM) in PSG employs the surface model described in section 4 to model the nucleus and the grains, while gas emissions are calculated employing a mono-layer non-LTE and LTE excitation model. The nucleus is assumed to be a spherical Lambertian emitting surface, while the dust is calculated as a diffuse and extended emitting component. The outgassing of dust-grains and parent molecular species is assumed to be isotropic and at constant expansion velocity, with photo-dissociation defining the lifetime and spatial extent of molecular species. Using these assumptions, the integrated number of molecules ($N$ [molecules]) within the FOV is defined as $N_{gas} = Q \cdot \tau \cdot f(x)$, where $\tau$ is the molecular lifetime [s], $f(x)$ is the filling-factor of the FOV with respect to the total coma, and $Q$ is the molecular production rate [molecules/s]. The definition and calculation of $f(x)$ is complex, with deriving an analytical form of $f(x)$ that employs Bessel functions for a circular FOV centered on the nucleus. Determining $f(x)$ for a non-centered FOV requires numerical Monte-carlo calculations, and for that purpose in PSG we compute the molecular $f(x)$ by interpolating from the tabulated $f(x)$ values derived by Xie & Mumma [87]. Two possible excitation regimes are considered in PSG for the molecules: non-LTE fluorescence (typically dominating the flux in the UV-Optical-IR range, [28–32]) and LTE excitation [70]. In both cases, PSG employs the molecular line-lists reported in section 3 for the computation of emission fluxes.

These assumptions are generally accurate enough (and widely employed by the community) to determine integrated column densities and molecular fluxes across the coma, yet the lifetime and velocity of the dust-grains can be mass/size/composition dependent and may differ from the surrounding gas environment. On the other hand, the strong relationship between visual magnitude (mainly defined by dust) and water production for 37 comets (see section 3.4, [50]) indicates that a common dust and gas outgassing scheme should be accurate enough for most cases, and it is the method employed by CEM. In CEM, we treat dust particles as behaving like the surrounding gas and it can be demonstrated that for a dust/gas mass ratio of 1.0, the brightness relationship determined by [50] is consistent with an average particle size of $r_{dust}$=3.4 microns (when assuming a dust particle density of $\rho$=0.5 kg/cm$^3$ and particle albedo as the nucleus of 0.04). When computing coma dust emission fluxes, we first derive the integrated water mass within the FOV as $M_{gas}$ [g] = $N_{gas} \cdot m_{gas} / A_g$, where $A_g$ is the Avogadro number 6.022E23 [molecules/mol] and $m_{gas}$ is the mean molar mass of the gas (18 [g/mol] for water), and then define the dust mass as $M_{dust} = M_{gas} \cdot DG$, where DG is the user-provided dust-to-gas mass ratio (1.0 is assumed to be typical, yet it is an adjustable parameter in CEM). The number of dust particles in the FOV is then $N_{dust} = M_{dust} / (4/3 \cdot A_{dust} \cdot r_{dust} \cdot \rho)$, where $A_{dust}$ is the particle cross-section ($\pi r_{dust}^2$), and the opacity due to dust is then $O_{dust} = N_{dust} \cdot (A_{dust}/A_{beam})$, where $A_{beam}$ is the area of the FOV at the comet. Ultimately, the effective emitting dust area is $A_{em} = A_{beam} \cdot (1.0 - \exp(-O_{dust}))$, which is the parameter employed to compute flux densities in PSG employing standard surface radiation terms.

## 5.3 Noise simulator

PSG currently includes an advanced sensitivity and noise calculator for different telescope and instrument configurations (e.g, coronagraph, interferometer), and for a diverse set of detector types (e.g, quantum, thermal, heterodyne). Computing noise for such a diverse set of modes over the whole electromagnetic spectrum range is complex and a first glance unattainable, yet when considering a set of constraints, background sources (see figure 6), guidelines and reasonable assumptions, the achieved accuracy can be very high. Importantly, when observing with ground-based observatories, PSG can also impose the effects of telluric absorption and noise on the synthetic spectra. The tool has access to a database of telluric transmittances pre-computed for 5 altitudes and 4 columns of water for each case pre-computed with PUMAS (see 5.1). The altitudes include that of Mauna-Kea/Hawaii (4200 m), Paranal/Chile (2600 m), SOFIA (14,000 m) and balloon observatories (35,000 m), while the water vapor column was established by scaling the tropical water profile by a factor of 0.1, 0.3 and 0.7 and 1. Opacities at 225 GHz, a typical metric to quantify water at radio wavelengths, can be estimated from the reported water column as $\tau_{225}$ = 0.0642 x PWV, where PWV is the amount of water in precipitable millimeters (reported by PSG).

With respect to the telescope configurations, PSG allows the user to define three types of telescope/instrument modes: a) single monolithic telescope, b) interferometric array, and c) a coronagraph instrument/telescope. In all cases, the integration of the fluxes is done over bounded and finite field-of-views and spectral ranges, with no spatial convolutions applied to the fields. The model for the coronagraph is relatively simple, yet the user can provide detailed contrast tables describing the performance of the instrument across $\lambda$/D. In the basic mode, PSG assumes that the throughput is minimum (1/contrast) within half the inner-working-angle (IWA), it reaches 50% at the IWA, and the throughput is maximum (100%) at 1.5 times the IWA.

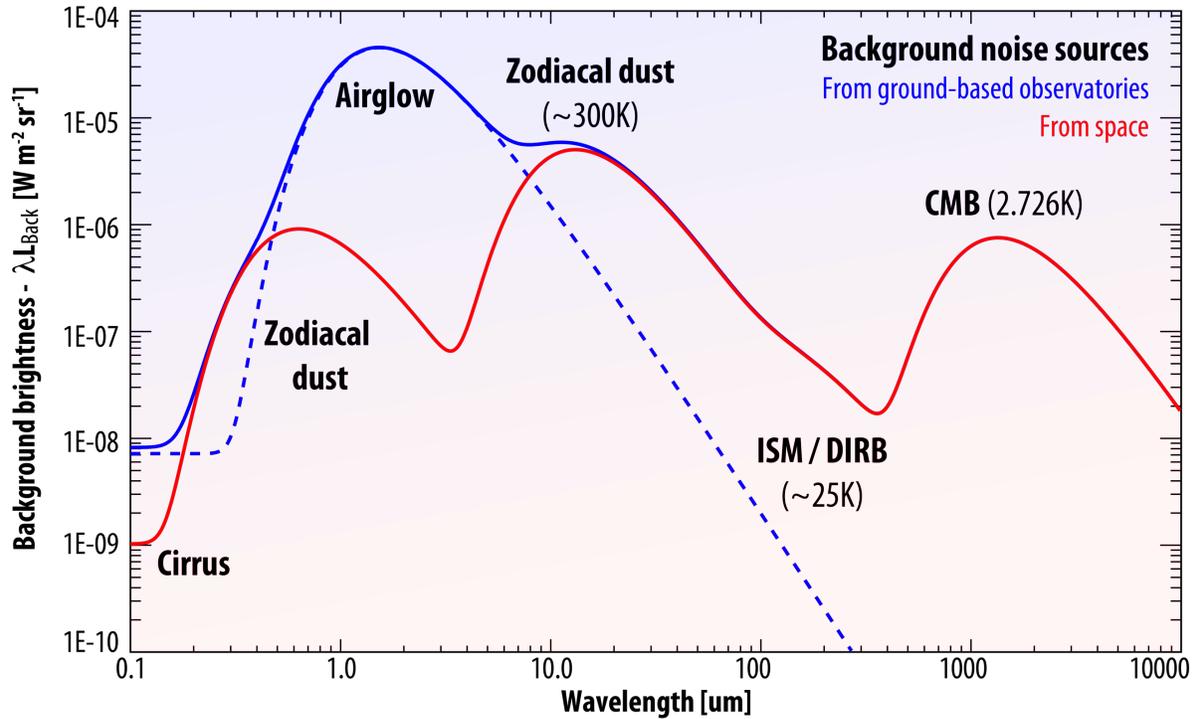

**Figure 6:** Background noise sources. When observing faint astronomical sources, the sensitivity is affected by the shot noise introduced by background and diffuse sources [88]. From space, the background is dominated by the faint and diffuse emission (thermal and scattered sunlight) from zodiacal dust, while airglow (a mixture of photoionization emissions, chemiluminescence and scattered sunlight) dominates the background for ground-based observations. Zodiacal dust fluxes depend greatly on the ecliptic longitude/latitude - in PSG, the noise simulator considers a scaling of 2 with respect to the minimum ecliptic pole values. PSG also employs a rudimentary yet relatively effective, approximation for atmospheric airglow.

Calculation of sensitivities requires a precise knowledge of the detector characteristics and the behavior of the signal and its noise under this regime. At short wavelengths (e.g., optical or near IR), the background photon counts follow a Poisson distribution, and the fluctuations are given by $\sqrt{N}$ where N is the mean number of photons received [89]. This Poisson distribution holds only in the case that the mean photon mode occupation number is small, n<<1. For a thermal background, the occupation number is given by the Bose-Einstein formula, so the opposite classical limit n>>1 is the usual situation at longer wavelengths for which hν<<kT. When n>>1, the photons do not arrive independently according to a Poisson process but instead are strongly bunched, and the fluctuations are of order N, instead of √N. In PSG, we employ Poisson √N statistics for the quantum (imager mode) and thermal detectors (NEP, D* modes), and the Dicke equation -which states that the noise is proportional to the background power rather than its square root- for radio/sub-millimeter detectors (TRX, receiver temperature mode). The formalism employed for the TRX module is based on the ALMA sensitivity calculator [90].

**Table 3. PSG noise formalism**

| Type of noise | Parameters | Detector specific noise equations |
|---|---|---|
| **TRX** Receiver temperature (radio) | $T_{RX}$ [K]: noise of receiver g: sideband (0:SSB, 1:DSB) $n_{pol}$: 1 (# of polarizations) $f_N$: 1 (# of baselines) **For interferometric systems** (e.g., ALMA): $n_{pol}$: 2 (assuming dual / full configuration) $f_N = n_{tele} \cdot (n_{tele}-1)$ | $L_{RJ} = 1E\text{-}18 \cdot \lambda^4 / 2kc$ $T_{source} = L \cdot L_{RJ}$ $T_{back} = L_{back}L_{RJ} + T_{ground}(1 - trn_{ground})$ $k_{sys} = (1+g)/(\eta_{Total}\ trn_{ground})$ $T_{sys} = k_{sys} [T_{RX} + \varepsilon_{optics}T_{optics} + T_{source} + T_{back}]$ [K] $f_\Omega = (\Omega_{Tele} / \Omega)$  Diffraction / FOV correction $d_v = 1E6 \cdot c \cdot d\lambda / \lambda^2$ $N_{total} = T_{sys} \cdot f_\Omega / \sqrt{(f_N \cdot n_{pol} \cdot d_v \cdot n_{exp} \cdot t_{exp})}$ [K] |
| **NEP** Noise Equivalent Power | NEP [W / √Hz]: sensitivity | $N_D = n_{pixels} \cdot n_{exp} \cdot t_{exp} \cdot (NEP \cdot \lambda \cdot 1E\text{-}6 / hc)^2$ [e$^{-2}$] |
| **D*** - Detectivity | D* [cm √Hz / W]: detectivity S [μm]: pixel size | $N_D = n_{pixels} \cdot n_{exp} \cdot t_{exp} \cdot ((S \cdot 1E\text{-}4 / D*) \cdot \lambda \cdot 1E\text{-}6/hc)^2$ |
| **Imager** – (e.g., CCD, CMOS, EMCCD, ICCD / MCP) | Read-noise [e- / pixel] Dark [e- / s / pixel] | $N_D = n_{pixels} \cdot n_{exp} \cdot [N_{read}^2 + (Dark \cdot t_{exp})]$ [e$^{-2}$] |

**The noise components with Poisson statistics (i.e., UV, optical, IR) are calculated as:**
$L_{e\text{-}} = \Omega \cdot A_{Tele} \cdot \eta_{eff} \cdot d\lambda \cdot t_{exp} \cdot n_{exp} \cdot 1E\text{-}6 / hc$   Radiance to detector electrons conversion factor
$N_{source} = L \cdot L_{e\text{-}}$   Noise introduced by the source itself [e$^{-2}$]
$N_{back} = (L_{back} + n_{ezo} \cdot L_{zodi}) \cdot L_{e\text{-}}$   Noise introduced by background sky sources [e$^{-2}$]
$N_{optics} = \varepsilon_{optics} \cdot L_{e\text{-}} \cdot (2E24 \cdot h \cdot c^2 / \lambda^5) / (exp(1E6 \cdot h \cdot c / (k \cdot T_{optics} \cdot \lambda)) - 1)$   Noise by the telescope [e$^{-2}$]
$N_{ground} = L_{e\text{-}} \cdot (1 - trn_{ground}) \cdot (2E24 \cdot h \cdot c^2 / \lambda^5) / (exp(1E6 \cdot h \cdot c / (k \cdot T_{ground} \cdot \lambda)) - 1)$   Noise for ground [e$^{-2}$]
$N_{Total} = \sqrt{(N_D + N_{source} + N_{back} + N_{optics} + N_{ground})}$   Total noise [e-]

**Parameters and constants:**
L [W / sr / m$^2$ / μm]: spectral radiance of the source
$L_{back}$ [W / sr / m$^2$ / μm]: spectral radiance of the background sources
$t_{exp}$ [s]: time per exposure
$n_{exp}$: total number of exposures
$n_{pixels}$: total number of pixels for $\Omega$ and $d\lambda$.
$n_{ezo}$: Exozodiacal dust scaler relative to Solar System zodiacal dust
$T_{optics}$ [K]: temperature of the optics
$\varepsilon_{optics}$: emissivity of the optics

$\eta_{\text{eff}}$: total throughput of the system (including quantum efficiencies)
$\Omega$ [steradian]: is the solid angle of the observations. It is wavelength dependent.
$A_{\text{Tele}}$ [m$^2$]: is the total collecting area of the observatory ($\eta_{\text{Tele}} \cdot \pi \cdot [D_{\text{Tele}}/2]^2$)
$\lambda$ [$\mu$m]: is the wavelength in microns
$\text{trn}_{\text{ground}}$: terrestrial transmittance
$T_{\text{ground}}$ [K]: temperature of the terrestrial atmosphere - 280
h [W s$^2$]: is Planck's constant - 6.6260693E-34
c [m / s]: is the speed of light - 299792458
k [J / K]: is Boltzmann's constant - 1.380658E-23

## 6. Online and retrieval capabilities

### 6.1 Web Interface and Application-Program-Interface (API)

The PSG tool can be accessed online at [https://psg.gsfc.nasa.gov](https://psg.gsfc.nasa.gov), where the user establishes the parameters of the simulation (discussed in the previous sections) and then performs a simulation request. The online presence is based on a PHP framework, which permits the user to input the parameters of the run, observe the 3D graphics of the orbital calculations, and plot the resulting spectra. The parameters entered by the user via the web GUI are stored in a configuration text file, which is then distributed among the spectroscopic PSG modules in order to perform the simulations. This configuration file can be downloaded and saved for future operations, and it is also internally saved on the PSG servers so every user always returns to her/his configuration file. The format of this file is in a relaxed form of XML (eXtensible Markup Language), the now-preferred file type across applications, while the resulting simulated spectra by PSG is provided in standard ASCII columns. Extensive documentation and tutorials on how to operate PSG are available at the site (see also Appendix A).

Importantly, PSG allows the user to perform operations remotely by employing a versatile Application Program Interface (API, see figure 7). The API operates by sending a configuration file to the PSG servers, which can be modified on the local machine as needed. Upon reception of the configuration file, PSG will compute the simulation and send back the planetary spectra. The main value of the API is that the user does not need to install / update the radiative transfer modules and databases on his/her computer - by performing a 'curl' command (via HTTPS), the user runs the simulations on high-performance NASA servers. Figure 7 explains the inner workings of the PSG modules, and how the user can enable / disable the different modules, and request for different spectral outputs.

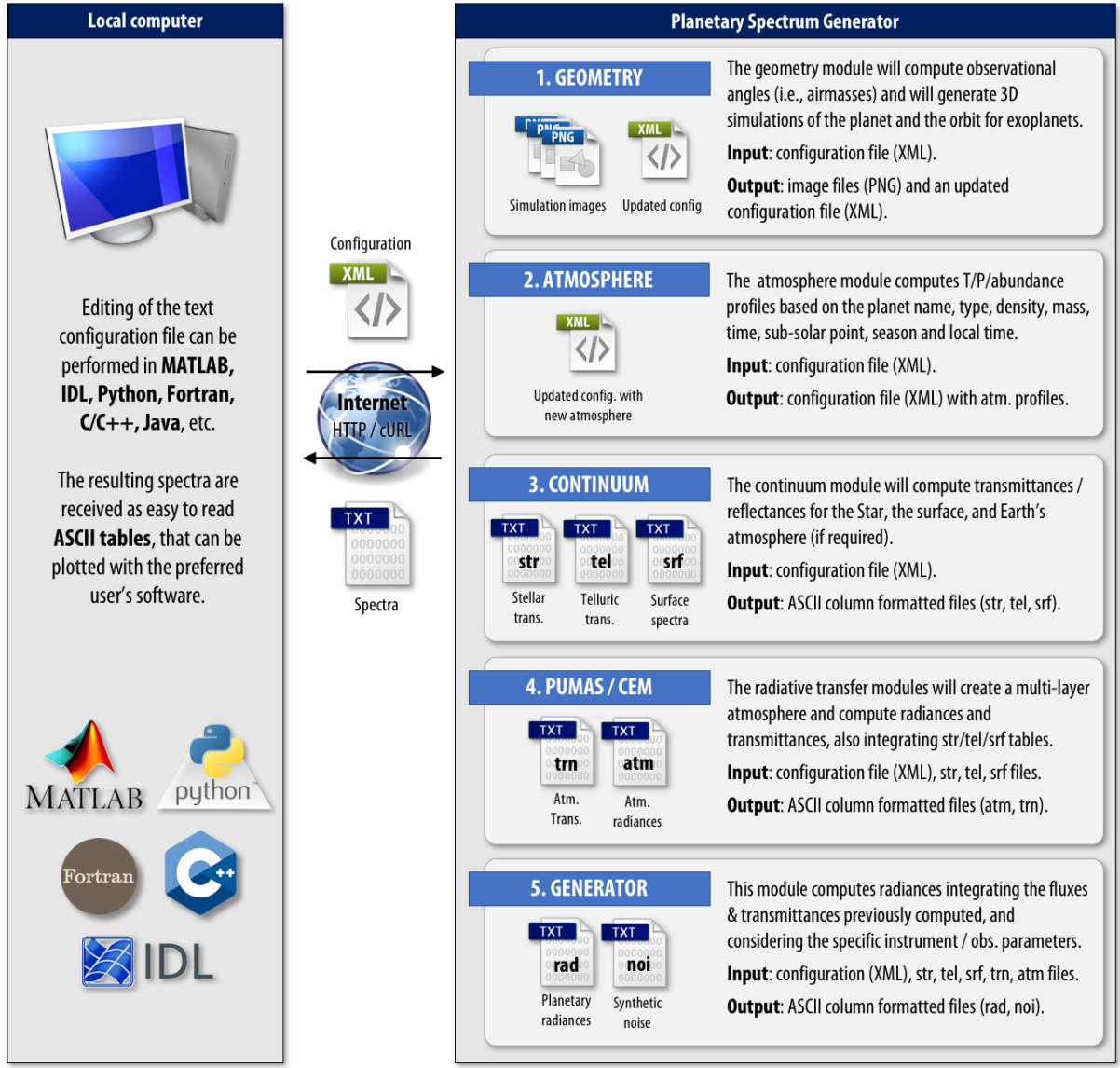

**Figure 7:** Application Program Interface. PSG can be called from a local machine via the 'curl' command that employs the HTTPS protocol for establishing the communication between the local machine and the PSG servers. The user sends a XML configuration file, and receives different type of spectral results (str, tel, srf, trn, atm, rad, noi), all standardized as text ASCII tables. The API will call the required modules (geometry, atmosphere, continuum, PUMAS/CEM, generator) in a sequential order, yet the user can also enable / disable modules as needed.

## 6.2 Retrievals

PSG permits the comparison of user-provided data to synthetically generated spectra and derives planetary parameters in the process (see examples in Figure 8). The retrieval process employs the Levenberg–Marquardt algorithm, also known as the damped least-squares method, in solving the non-linear least squares problem. This method interpolates between the Gauss–Newton algorithm and the method of gradient descent, with the Levenberg–Marquardt method being more robust, meaning that it finds a solution even if the initial conditions are far from the final solution. The Levenberg-Marquardt algorithm is based on the MPFIT program [91], which draws from the robust package called MINPACK-1.

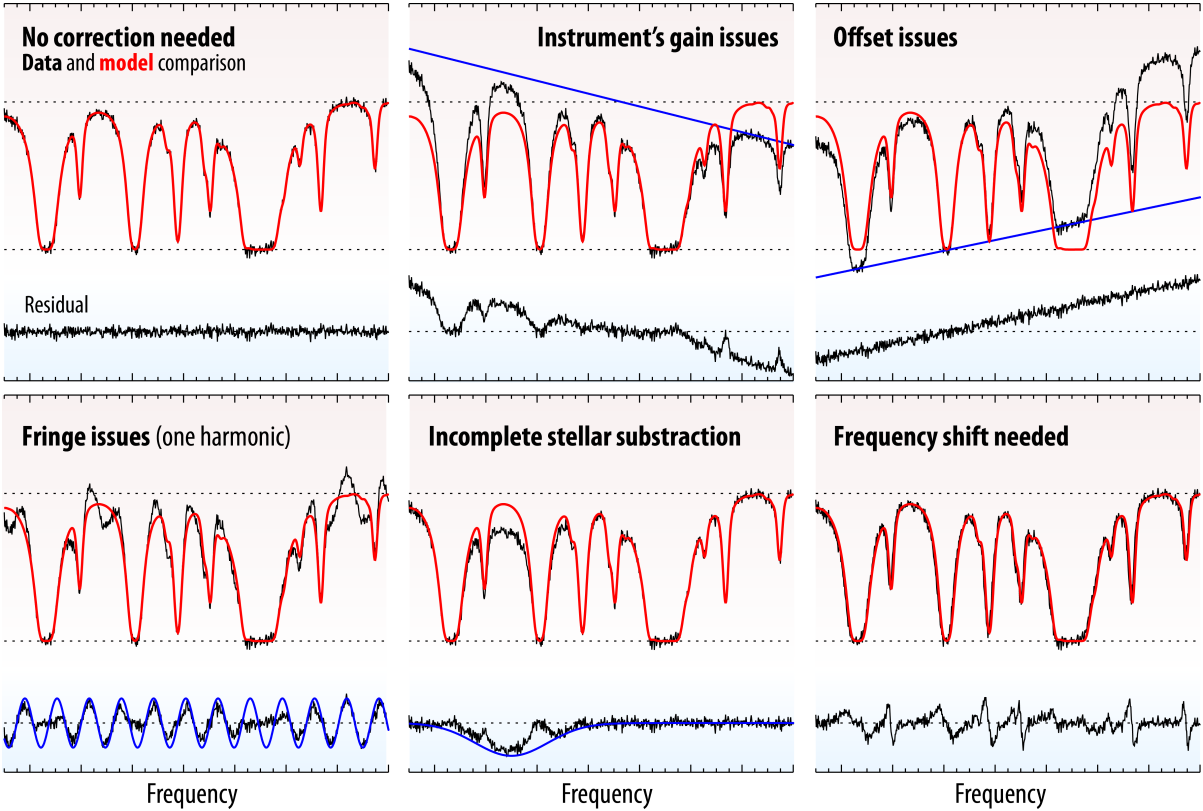

**Figure 8:** Beyond performing the retrieval of planetary parameters, PSG can also correct for several typical issues affecting spectroscopic data. The different panels show how these issues could affect the data, and their impact on the residuals. The methods employed to remove these instrumental effects are described in [92].

The user-provided file should be formatted as a text file with two or three columns. The first column should indicate the frequency/wavelength of each pixel, while the second column should describe the measured flux. A third column (optional) should indicate the 1-sigma uncertainty in flux, and if no error is provided, PSG will assume a 5% uncertainty. The intensity of the data can be scaled to a flux physical unit by a scaling factor, while the user can select from a broad range of frequency/wavelength units (e.g., GHz, cm⁻¹, μm) and flux units (e.g., W/m²/μm, Jy). Beyond fitting the data to the model, PSG can also correct for a broad range of issues typically affecting spectroscopic data (see figure 8). By employing pre-computed telluric spectra (see section 5.3), PSG can also fit the water and column parameters affecting ground-based data.

The uncertainty in the retrieved parameters is computed from the covariance matrix, which is in turn calculated from the Jacobian matrix at convergence. This is the standard statistical method used to compute parameter errors from Levenberg-Marquardt retrievals. In order to compensate for the quality of the fit, the reported uncertainties are scaled by the square root of the reduced chi-square. Since the retrieval method involves running dozens of spectroscopic simulations, a retrieval can take up to a couple of minutes for intensive scattering radiative-transfer calculations. It should take only a few seconds when running emission and cometary retrievals, since these computations are based on a heavily optimized radiative transfer package, which employs pre-computed non-LTE fluorescence efficiencies.

## 7.0 Current limitations of PSG and future steps

As discussed in the previous sections, PSG can be used to synthesize a wide range of planetary fluxes under a variety of observing conditions, yet there are particular cases and modes not yet supported by PSG. We are actively working on addressing these, and we expect that these additional capabilities will become available to the community in the next years. We do think that the main analytical and numerical methods associated with the radiative-transfer calculations (e.g., scattering, layer-by-layer, line-by-line, correlated-k, surface synthesis) have been already addressed, with much of the improvements happening in the input parameters for these models.

**Atomic / ionized species:** as presented in Table 1, PSG can synthesize spectra for hundreds of molecules, yet for atomic species, only a handful are sparsely supported (e.g., Na, K, O, C). Numerous spectroscopic databases for atomic species do exist (e.g., VAMDC, NIST, Kurucz), yet they rarely provide information about line broadening and properly modeling these emissions / absorptions requires of ad-hoc radiative-transfer analysis. Furthermore, these species are typically present in non-equilibrated environments (e.g., non-LTE, ionized), and the excitation process has to be properly tailored for each regime. Our emphasis in the near future will be in collecting and acquiring reliable broadening information for the most commonly observed atoms, that we will then use to compute line-by-line spectra and cross-sections.

**Lineshape broadening and high-temperature excitation:** most spectroscopic information regarding lineshapes has been done for an air mixture ($N_2+O_2$) as collisional partner, and it is the information provided by the main molecular repositories (e.g., HITRAN and GEISA), yet HITRAN does provide lineshape information for non-terrestrial atmospheres for a selection of species. Other databases (e.g., JPL and CDMS) do not provide lineshape information. Extrapolating these broadening coefficients to other collisional regimes (e.g., $CO_2$, $H_2$, He) is far from trivial. Some approximations and numerical methods do exist to perform this transformation (e.g., Robert Bonamy complex mechanism), yet such methods have to be tailored and parameterized using empirical information. In addition, most databases (e.g., HITRAN, GEISA, JPL, CDMS) have been designed to provide spectroscopic information for lines which are active at "terrestrial" temperatures (~300K), and therefore they are relatively incomplete for computing high-temperature spectra (T>1500K). In the last years, the ExoMol spectroscopic database has started compiling information about line broadening parameters for certain species in a $H_2$/He atmosphere and for high temperatures. They have also recently released a numerical method (ExoCross) to compute cross-sections using a variety of spectral databases [93]. We are actively working with these teams (e.g., HITRAN, ExoMol), and as new spectroscopic information is released, we will integrate them into PSG.

**Non-equilibrated regimes:** in instances where collisions no longer establish the radiative equilibrium of the molecules, or chemistry leads to non-equilibrated states, the atmosphere can be considered to be in non-thermodynamic equilibrium. Such disequilibrium can be thermal, mechanical, chemical, and/or radiative, and it is for instance the case of exospheres or highly radiated environments. Cometary atmospheres are an example of non-LTE, as the low densities (low collisional rates) and high solar influx lead to radiative disequilibrium, as demonstrated by the strong fluorescence emissions dominating their infrared spectra. In PSG, we model this disequilibrium by ingesting fluorescence efficiencies (g-factors) which are pre-computed for a particular rotational temperature and solar radiation flux. These linelists can be applied to other similar exospheres of comparable low-collissional rate and high-insolation rate. We expect in the future to integrate other non-LTE processes in PSG, by providing specifically tailored linelists and population schemes for each regime.

## 8.0 Conclusions

We have developed a tool for planetary spectroscopy (Planetary Spectrum Generator [PSG]), in which several areas of research, such as orbital dynamics, molecular and laboratory spectroscopy, quantum mechanical modeling, radiative transfer and instrument modeling, are integrated into a powerful and realistic synthesis method, enabling the retrieval of planetary parameters from remote sensing data, efficient planning of mission strategies, interpretation of current and future planetary data, calibration of spectroscopic data and development of new instrument/spacecraft concepts. The tool relies on validated methodologies and comprehensive spectroscopic repositories, leading to highly accurate and realistic results. More importantly, the tool is publicly available with a user-friendly graphical interface (even accessible from mobile devices), and it does not require the complex installations and fail-prone compilation procedures typically required when operating scientific packages. The current methods allow PSG to accurately compute fluxes, transmittances, reflectances and emissivities for a wide range of planetary objects (e.g., terrestrial planets, gas giants, asteroids, comets, icy moons, TNOs, KBOs, exoplanets), for a broad range of wavelengths (UV/Vis/near-IR/IR/far-IR/THz/sub-mm/Radio) from any observatory (e.g., JWST, ALMA, Keck, SOFIA), any orbiter (e.g., MRO, ExoMars, Cassini, New Horizons), or any lander (e.g., MSL). The diversity in potential applications for PSG is shown in figures 9, in which high-resolution spectra of Mars and of comet Boattini are fitted to PSG models, and used to retrieve molecular abundances on these bodies. More importantly, the tool can serve a major role in the development and design of new instrumentation and observational strategies, as demonstrated by the exoplanet simulations presented in figure 10 of a transit JWST exoplanet spectra, and the coronagraphic observations of a terrestrial planet with the concept LUVOIR observatory.

Because of its modular architecture, PSG is more than a tool - it is a framework for planetary spectroscopy, permitting each module to mature and increase in sophistication over time. Some of the areas in which we expect more growth and new developments are: 1) integration of several molecular and atomic linelists applicable to a broader range of excitation and collisional regimes, 2) integration of new instrument models applicable to a broader range of telescope / instrument configurations, and 3) development of climatological and chemical modules in order to further serve the diversity of exoplanetary atmospheres. These and other advances will greatly serve the planetary community in general, ultimately enabling to address of broad range of scientific investigations.

**Acknowledgements**

GLV acknowledges support from the NASA's Exoplanets Research Program (16-XRP16_2-0071), NASA's Emerging Worlds Research Program (15-EW15_2-0175) and NASA's Goddard Fellows Innovation Challenge (GFIC-2017). AMM acknowledges support from 16-XRP16_2-007. We thank Drs. Tilak Hewagama, Michael Wolff, Michael Mumma, Emily Wilson and Giuliano Liuzzi for their helpful guidance and advice.

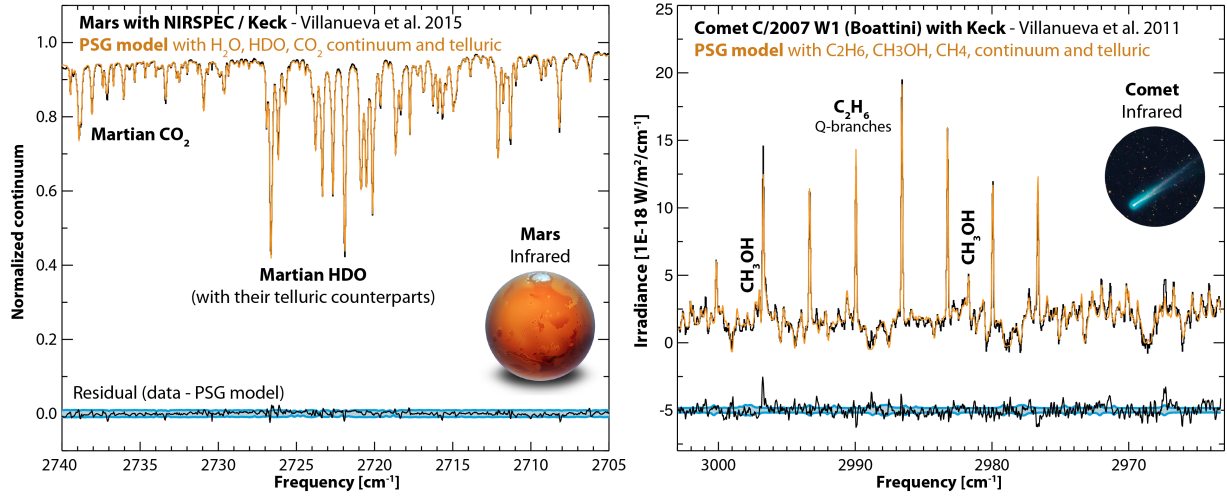

**Figure 9:** Application of PSG to retrieve molecular abundances from high-resolution spectra of Mars and of comet Boattini. **Left:** Comparison between high-resolution of HDO and CO₂ on Mars obtained with NIRSPEC/Keck-II [1], and PSG simulations that also include correction for telluric absorption and instrumental fringing. **Right**: Comparison between ground-based observations of comet C/2007 W1 (Boattini) with NIRSPEC/Keck-II [32] and simulations including several organic species and dust/nucleus emission

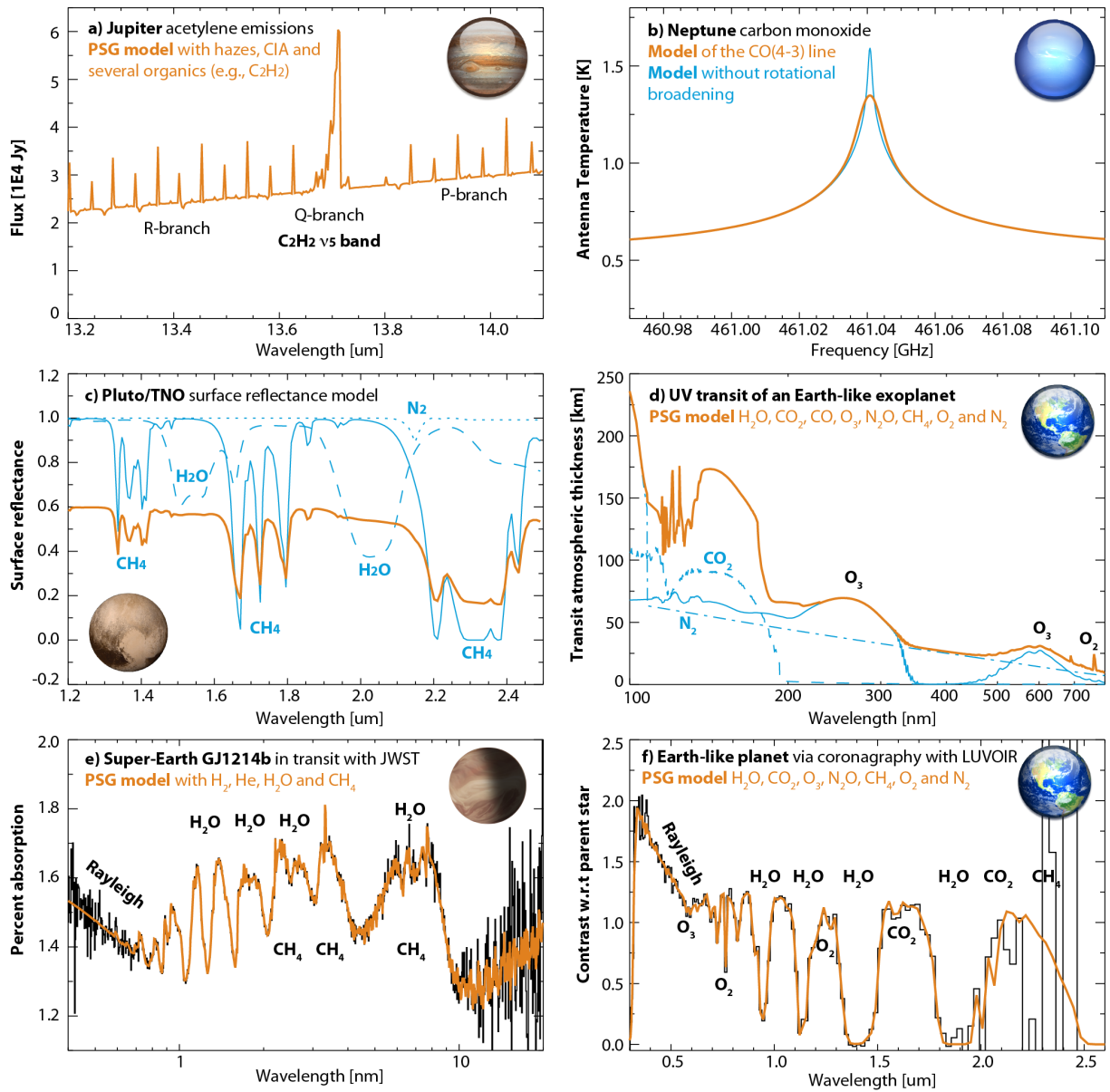

**Figure 10:** Example of simulations performed with PSG on a diversity of planetary bodies. a) Simulations of the Jovian acetylene ($C_2H_2$) emissions in the mid-IR as observed with a space observatory (compare to [94]); b) Simulations of the Neptune carbon-monoxide (CO) J=5-4 rotational line with a sub-millimeter observatory (compare to [95]); c) Surface reflectance modeling of a Pluto-like object as observed at near-IR wavelengths (compare to [58] ); d) Ultraviolet transit spectrum of an Earth-like planet containing the signatures of several key species in our atmospheres (compare to [96]); e) Simulated transit spectra of super-Earth GJ-1214b as observed with JWST (compare to [33]). f) Simulated coronagraphy spectra of an Earth-like planet as observed with LUVOIR (compare to [97]).

# Appendix A

The modules in the Planetary Spectrum Generator (PSG) operate by establishing their parameters by reading the information from a configuration file (associated to each user). The format of the file is a relaxed form of XML (eXtensible Markup Language), the now preferred file type across applications. Every parameter is entered as an individual line in the form of "<OBJECT>Mars". For parameters allowing multiple values, these are entered separated by commas (CSV) in the form of "<ATMOSPHERE-GAS>H2O,CH4,CO".

The user can modify these parameters as needed (within the allowed ranges, the values in brackets for the "Text" type indicate maximum number of characters), and some modules will also populate and modify these parameters as needed. For instance, the "GEOMETRY" module of PSG, will consider information about the date, object and observing geometry and will calculate observing angles and other geometric parameters that will be stored in the configuration file. The user is also free to override these calculations by the GEOMETRY module and to consider another set of geometrical parameters.

| Keyword | Range | Description |
|---|---|---|
| **OBJECT** | | |
| OBJECT | Text(50) | Object type (e.g., Exoplanet) or object name for the main bodies in the Solar System |
| OBJECT-NAME | Text(50) | Object name |
| OBJECT-DATE | Text(20) | Date of the observation (yyyy/mm/dd hh:mm) in Universal time [UT] |
| OBJECT-DIAMETER | 1E-2 to 1E8 | Diameter of the object [km] |
| OBJECT-GRAVITY | 0 to 1E35 | Gravity/density/mass of the object |
| OBJECT-GRAVITY-UNIT | Text(10) | Unit for the OBJECT-GRAVITY field, g:'Surface gravity [m/s2]', rho:'Mean density [g/cm3]', or kg:'Total mass [kg]' |
| OBJECT-STAR-DISTANCE | 0 to 1E5 | Distance of the planet to its parent star [AU] |
| OBJECT-STAR-VELOCITY | -1E4 to 1E4 | Velocity of the planet with respect to its parent star (in the observer-planet vector) [km/s] |
| OBJECT-SOLAR-LONGITUDE | -360 to 360 | Sub-solar east longitude [degrees] / For exoplanets, longitude of periapse [degrees] |
| OBJECT-SOLAR-LATITUDE | -90 to 90 | Sub-solar latitude [degrees] / For exoplanets, orbital eccentricity |
| OBJECT-SEASON | 0 to 360 | Angular parameter (season/phase) that defines the position of the planet moving along its Keplerian orbit, where 0:'Northern Hemisphere Spring Equinox', 90:'Northern Hemisphere Summer Solstice', 180:'Northern Hemisphere Autumn Equinox', 270:'Northern hemisphere Winter Solstice' [degrees] |
| OBJECT-STAR-TYPE | Text(1) | Stellar type of the parent star [O/B/A/F/G/K/M] |
| OBJECT-STAR-TEMPERATURE | 1 to 1E5 | Temperature of the parent star [K] |
| OBJECT-STAR-RADIUS | 1E-3 to 1E8 | Radius of the parent star [Rsun] |
| OBJECT-OBS-LONGITUDE | -360 to 360 | Sub-observer east longitude [degrees] |
| OBJECT-OBS-LATITUDE | -90 to 90 | Sub-observer latitude [degrees] |
| OBJECT-OBS-VELOCITY | -1E4 to 1E4 | Relative velocity between the observer and the object [km/s] |
| OBJECT-PERIOD | 0 to 1E8 | This field is computed by the geometry module - It is the apparent rotational period of the object as seen from the observer [Earth days] |
| **GEOMETRY** | | |
| GEOMETRY | Text(20) | Type of observing geometry, Telescope / Nadir / Limb / Solar / Stellar / Lookingup / LookingSolar |
| GEOMETRY-REF | Text(50) | Reference geometry (e.g., TGO, Maunakea), default is user defined or 'User' |
| GEOMETRY-OFFSET-NS | -1E6 to 1E6 | Horizonal offset (w.r.t to the Equator) of the sub-observer location |
| GEOMETRY-OFFSET-EW | -1E6 to 1E6 | Vertical offset (w.r.t to the Equator) of the sub-observer location |
| GEOMETRY-OFFSET-UNIT | Text(10) | Unit of the GEOMETRY-OFFSET field, arcsec / arcmin / degree / km / diameter |
| GEOMETRY-OBS-ALTITUDE | 0 to 1E8 | Distance between the observer and the surface of the planet |
| GEOMETRY-ALTITUDE-UNIT | Text(10) | Unit of the GEOMETRY-OBS-ALTITUDE field, AU / km / diameter and pc:'parsec' |
| GEOMETRY-USER-PARAM | 0 to 1000 | Parameter for the selected geometry, for Nadir / Lookingup this field indicates the zenith angle [degrees], for limb / occultations this field indicates the atmospheric height [km] being sampled |
| GEOMETRY-STELLAR-TYPE | Text(1) | For stellar occultations, this field indicates the type of the occultation star [O/B/A/F/G/K/M] |
| GEOMETRY-STELLAR-TEMPERATURE | 1 to 1E5 | For stellar occultations, this field indicates the temperature [K] of the occultation star |
| GEOMETRY-STELLAR-MAGNITUDE | -30 to 30 | For stellar occultations, this field indicates the brightness [magnitude] of the occultation star |
| GEOMETRY-OBS-ANGLE | 0 to 360 | This field is computed by the geometry module - It indicates the angle between the observer and the planetary surface |
| GEOMETRY-SOLAR-ANGLE | 0 to 360 | This field is computed by the geometry module - It indicates the angle between the Sun and the planetary surface |
| GEOMETRY-PHASE | -360 to 360 | This field is computed by the geometry module - It indicates the phase between the Sun and observer |
| GEOMETRY-PLANET-FRACTION | 0 to 1 | This field is computed by the geometry module - It indicates how much the beam fills the planetary area (1:maximum) |
| GEOMETRY-STAR-FRACTION | 0 to 1 | This field is computed by the geometry module - It indicates how much the beam fills the parent star (1:maximum) |
| GEOMETRY-STAR-DISTANCE | -1 to 1E8 | This field is computed by the geometry module - It indicates the projected distance between the beam and the parent star |
| GEOMETRY-ROTATION | CSV values | This field is computed by the geometry module - It indicates the rotational Doppler shift [km/s] affecting the spectra and the spread of rotational velocities [km/s] within the FOV |
| **ATMOSPHERE** | | |
| ATMOSPHERE-STRUCTURE | Text(100) | The structure of the atmosphere, None / Equilibrium:'Hydrostatic equilibrium' / Coma:'Cometary expanding coma' |
| ATMOSPHERE-PRESSURE | 0 to 1E35 | For equilibrium atmospheres, this field defines the surface pressure; while for cometary coma, this field indicates the gas production rate |
| ATMOSPHERE-PUNIT | Text(20) | The unit of the ATMOSPHERE-PRESSURE field, Pa:Pascal / bar / kbar / mbar / ubar / at / atm / torr / psi / gas:'molecules / second' / gasau:'molecules / second at rh=1AU' |
| ATMOSPHERE-TEMPERATURE | 1 to 1E4 | For atmospheres without a defined P/T profile, this field indicates the temperature across all altitudes |
| ATMOSPHERE-WEIGHT | 1 to 1E3 | Molecular weight of the atmosphere [g/mol] |
| ATMOSPHERE-NGAS | 0 to 20 | Number of gases to include in the simulation |
| ATMOSPHERE-GAS | CSV text | Name of the gases to include in the simulation, e.g 'H2O,CO2' |
| ATMOSPHERE-TYPE | CSV text | Sub-type of the gases, e.g. 'HIT[1],HIT[2]' |
| ATMOSPHERE-ABUN | CSV values | Abundance of gases. The values can be assumed to be same across all altitudes/layers [%,ppmv,ppbv,pptv,m-2], or as a multiplier [scaler] to the provided vertical profile (see ATMOSPHERE-LAYERS-MOLECULES) |
| ATMOSPHERE-UNIT | CSV text | Unit of the ATMOSPHERE-ABUN field, % / ppmv / ppbv / pptv / m2:'molecules/m2' / scl:'scaler of profile' |
| ATMOSPHERE-TAU | CSV values | For expanding cometary coma, this field indicates the photodissociation lifetime of the molecules [s] |
| ATMOSPHERE-NAERO | 0 to 20 | Number of aerosols to include in the simulation |
| ATMOSPHERE-AEROS | CSV text | Name of the aerosols to include in the simulation |
| ATMOSPHERE-ATYPE | CSV text | Sub-type of the aerosols |
| ATMOSPHERE-AABUN | CSV values | Abundance of aerosols. The values can be assumed to be same across all altitudes/layers [%,ppm,ppb,ppt,Kg/m2], or as a multiplier [scaler] to the provided vertical profile (see ATMOSPHERE-LAYERS-MOLECULES) |
| ATMOSPHERE-AUNIT | CSV text | Unit of the ATMOSPHERE-AABUN field, % / ppmv / ppbv / pptv / m2:'molecules/m2' / scl:'scaler of profile' |
| ATMOSPHERE-ASIZE | CSV values | Effective radius of the aerosol particles [um] |
| ATMOSPHERE-NMAX | 0 to 100 | When performing scattering aerosols calculations, this parameter indicates the number of n-stream pairs - Use 0 for extinction calculations |
| ATMOSPHERE-LMAX | 0 to 100 | When performing scattering aerosols calculations, this parameter indicates the number of scattering Legendre polynomials used for describing the phase function - Use 0 for extinction calculations |
| ATMOSPHERE-DESCRIPTION | Text(200) | Description establishing the source/reference for the vertical profile |
| ATMOSPHERE-LAYERS | 0 to 999 | Number of layers of the atmospheric vertical profile |
| ATMOSPHERE-LAYERS-MOLECULES | CSV text | Molecules quantified by the vertical profile |
| ATMOSPHERE-LAYER | CSV values | Values for that specific layer, Pressure[bar] / Temperature[K] / gases[mol/mol] / aerosols [kg/kg] |

| SURFACE | | |
|---|---|---|
| SURFACE-TEMPERATURE | 0 to 1E5 | Temperature of the surface [K] |
| SURFACE-ALBEDO | 0 to 1.0 | Albedo the surface [0:non-reflectance, 1:fully-reflective] |
| SURFACE-EMISSIVITY | 0 to 1.0 | Emissivity of the surface [0:non-emitting, 1:perfect-emitter] |
| SURFACE-PHASEG | -1 to 1 | One-term Henyey-Greenstein g-factor [0:isotropic, -1:backward scatterer, +1:forward scatterer] |
| SURFACE-GAS-RATIO | 0 to 1E3 | For expanding cometary coma, this value indicates an scaling value for the dust in the coma [1:typical dust/gas ratio] |
| SURFACE-NSURF | 0 to 20 | Number of components describing the surface properties [areal mixing] |
| SURFACE-SURF | CSV text | Name of the surface components to be included in the simulation |
| SURFACE-TYPE | CSV text | Sub-type of the surface components |
| SURFACE-ABUN | CSV values | Relative abundance of the surface components |
| SURFACE-UNIT | CSV text | Unit of the SURFACE-ABUN field, % / ppm / ppv |
| SURFACE-THICK | CSV values | Thickness for each surface component [um] |
| GENERATOR | | |
| GENERATOR-RANGE1 | 1E-5 to 1E7 | Lower spectral range for the simulation |
| GENERATOR-RANGE2 | 1E-5 to 1E7 | Upper spectral range for the simulation |
| GENERATOR-RANGEUNIT | Text(10) | Unit of the GENERATOR-RANGE fields, um / nm / mm / An:'Angstrom' / cm:'Wavenumber [cm-1]' / MHz / GHz / kHz |
| GENERATOR-RESOLUTION | 1E-6 to 1E8 | Spectral resolution for the simulation. PSG assumes that the sampling resolution is equal is to the instrumental resolution, yet radiative transfer resolutions are always performed at the necessary/higher resolutions in order to accurately describe the lineshapes |
| GENERATOR-RESOLUTIONUNIT | Text(10) | Unit of the GENERATOR-RESOLUTION field, RP:'Resolving power' / um / nm / mm / An:'Angstrom' / cm:'Wavenumber [cm-1]' / MHz / GHz / kHz |
| GENERATOR-GAS-MODEL | Text(1) | Flag indicating whether to include molecular signatures as generated with PUMAS or CEM [Y/N] |
| GENERATOR-CONT-MODEL | Text(1) | Flag indicating whether to include continuum signatures as generated by the surface, the star (when in the field) and dust/nucleus (when synthesizing comets) [Y/N] |
| GENERATOR-CONT-STELLAR | Text(1) | Flag indicating whether to include stellar absorption signatures in the reflected sunlight / stellar spectra [Y/N] |
| GENERATOR-TRANS-SHOW | Text(1) | Flag indicating whether we are synthesizing planetary spectra (not of Earth) as observed with a ground-based telescope. This flag will ensure that the noise module properly includes telluric signatures. |
| GENERATOR-TRANS-APPLY | Text(1) | Flag indicating whether to show the spectra divided by the telluric transmittance [N], or as observed and affected by transmittance [Y] |
| GENERATOR-TRANS | Text(20) | Keyword [SS-WW] indicating the site [SS] and water abundance [WW]. Values of SS are 00:'0m (sea level)', 01:'2,600m (8,500 feet)', 02:'4,200m (14,000 feet)', 03:'14,000m (46,000 feet)', 04:'35,000m (120,000 feet)'. Values of WW are 00:'10% tropical', 01:'30% tropical', 02:'70% tropical', 03:'100% tropical' |
| GENERATOR-RADUNITS | Text(20) | Radiation unit for the generated spectra (e.g., Jy) |
| GENERATOR-LOGRAD | Text(1) | Flag indicating whether to show the spectra employing a logarithmic scale |
| GENERATOR-TELESCOPE | Text(10) | Type of telescope, SINGLE:'single dish telescope / instrument' / ARRAY:'Interferometric array' / CORONA:'Coronagraph' |
| GENERATOR-BEAM | 1E-3 to 1E6 | Full width half-maximum (FWHM) of the instrument's beam |
| GENERATOR-BEAM-UNIT | Text(20) | Unit of the GENERATOR-BEAM field, arcsec / arcmin / degree / km / diameter:'beamsize in terms of the planet's diameter' / diffrac:'defined by the telescope diameter and center wavelength' |
| GENERATOR-DIAMTELE | 1E-5 to 1E5 | Diameter of the main reflecting surface of the telescope / instrument [m] |
| GENERATOR-TELESCOPE1 | 1E-20 to 1000 | For interferometers, the number of telescopes; for coronagraphs, the instrument's contrast |
| GENERATOR-TELESCOPE2 | 0 to 1E20 | For coronographic observations, the exozodi level w.r.t to the terrestrial level |
| GENERATOR-TELESCOPE3 | 0.1 to 100 | For coronographic observations, the inner working angle (IWA) in units of [L/D] |
| GENERATOR-NOISE | Text(20) | Keyword identifying the noise model to consider, NO:'None' / TRX:'Receiver temperature / radio' / RMS:'Constant noise in radiation units' / BKG:'Constant noise with added background' / NEP:'Power equivalent noise detector model' / D*:'Detectability noise detector model' / CCD:'Image sensor' |
| GENERATOR-NOISETIME | 0 to 1E7 | Exposure time per frame [sec] |
| GENERATOR-NOISEFRAMES | 1 to 1E5 | Number of exposures |
| GENERATOR-NOISEPIXELS | 1 to 1E9 | Total number of pixels that encompass the beam (GENERATOR-BEAM) and the spectral unit (GENERATOR-RESOLUTION) |
| GENERATOR-NOISE1 | 0 to 1E20 | First noise model parameter - For RMS, 1-sigma noise; for TRX, the receiver temperature; for BKG, the 1-sigma noise; for NEP, the sensitivity in W/sqrt(Hz); for DET, the sensitivity in cm.sqrt(Hz)/W; for CCD, the read noise [e-] |
| GENERATOR-NOISE2 | 0 to 1E20 | Second noise model parameter - For RMS, not used; for TRX, the sideband g-factor; for BKG, the not used; for NEP, not used; for DET, the pixel size [um]; for CCD, the dark rate [e-/s] |
| GENERATOR-NOISEOEFF | 0 to 1 | Total throughput of the telescope+instrument, from photons arriving to the main mirror to photons being quantified by the detector [0:none to 1:perfect]. The user can provide wavelength dependent values as neff@wavelength[um] (e.g., '0.087@2.28,0.089@2.30,0.092@2.31,0.094@2.32,...') |
| GENERATOR-NOISEOEMIS | 0 to 1 | Emissivity of the telescope+instrument optics [0 to 1] |
| GENERATOR-NOISEOTEMP | 0 to 1E4 | Temperature of the telescope+instrument optics [K] |
| GENERATOR-INSTRUMENT | Text(500) | Text describing if an instrument template was used to define the GENERATOR parameters |
| **RETRIEVAL** | | |
| RETRIEVAL-RANGEUNIT | Text(10) | Spectral unit of the user-provided data for the retrieval, um / nm / mm / An:'Angstrom' / cm:'Wavenumber [cm-1]' / MHz / GHz / kHz |
| RETRIEVAL-RESOLUTION | 1E-6 to 1E8 | Instrument's spectral resolution [FWHM] of the user-provided data. This value is independent of the sampling rate of the data, and refers to the actual spectral resolution of the instrument. |
| RETRIEVAL-RESOLUTIONUNIT | Text(10) | Unit of the RETRIEVAL-RESOLUTION field, RP:'Resolving power' / um / nm / mm / An:'Angstrom' / cm:'Wavenumber [cm-1]' / MHz / GHz / kHz |
| RETRIEVAL-FLUXSCALER | 0 to 1E30 | Scaling value to be applied to all fluxes of the user-provided data file. |
| RETRIEVAL-FREQSHIFT | -1E5 to 1E5 | Frequency/wavelength shift to be applied to the data. |
| RETRIEVAL-FLUXLABELS | Text(100) | Labels for the columns of the data file. |
| RETRIEVAL-FITGAIN | -1 to 2 | Polynomial degree of the instrument's gain function, -1:None, 0:Constant, 1:Sloped, 2:Quadratic, etc. |
| RETRIEVAL-REMOVEOFFSET | -1 to 2 | Polynomial degree of the residual offset, -1:None, 0:Constant, 1:Sloped, 2:Quadratic, etc. |
| RETRIEVAL-REMOVEFRINGE | 0 to 9 | Maximum number of spectral fringes to be removed from the data. |
| RETRIEVAL-FITSTELLAR | Text(1) | Flag indicating whether to fit the intensity of the solar/stellar features [Y/N] |
| RETRIEVAL-FITFREQ | Text(1) | Flag indicating whether to refine the spectral calibration [Y/N] |
| RETRIEVAL-FITRESOLUTION | Text(1) | Flag indicating whether to fit the spectral resolution [Y/N] |
| RETRIEVAL-FITTELLURIC | Text(1) | Flag indicating whether to refine the telluric features [Y/N]. This is done by perturbing the selected telluric column/water abundances. |
| RETRIEVAL-NVARS | 0 to 9 | Number of model variables to be retrieved. |
| RETRIEVAL-VARIABLES | CSV text | Name of the variables of the retrieval (comma separated). |
| RETRIEVAL-VALUES | CSV text | A-priori values of the retrieval parameters (comma separated). |
| RETRIEVAL-MIN | CSV text | Lower boundary permitted for each parameter (comma separated). |
| RETRIEVAL-MAX | CSV text | Upper boundary permitted for each parameter (comma separated). |
| RETRIEVAL-UNITS | CSV text | Magnitude unit of the a-priori and boundary entries (comma separated). |
| RETRIEVAL-STATUS | Text(30) | Flag indicating the status of the retrieval suite (e.g., RUNNING, OK) |

## References

[1] Villanueva GL, Mumma MJ, Novak RE, Käufl HU, Hartogh P, Encrenaz T, et al. Strong water isotopic anomalies in the martian atmosphere: Probing current and ancient reservoirs. Science 2015;348:218–221. doi:10.1126/science.aaa3630.

[2] Hartogh P, Lis DC, Bockelée-Morvan D, de Val-Borro M, Biver N, Küppers M, et al. Ocean-like water in the Jupiter-family comet 103P/Hartley 2 2011;478:218–220. doi:10.1038/nature10519.

[3] Villanueva GL, Mumma MJ, Bonev BP, Di Santi MA, Gibb EL, Böhnhardt H, et al. A Sensitive Search for Deuterated Water in Comet 8P/Tuttle. The Astrophysical Journal Letters 2009;690:L5–L9. doi:10.1088/0004-637X/690/1/L5.

[4] Stern SA, Bagenal F, Ennico K, Gladstone GR, Grundy WM, McKinnon WB, et al. The Pluto system: Initial results from its exploration by New Horizons. Science 2015;350:aad1815. doi:10.1126/science.aad1815.

[5] Grundy WM, Binzel RP, Buratti BJ, Cook JC, Cruikshank DP, Dalle Ore CM, et al. Surface compositions across Pluto and Charon. Science 2016;351:aad9189–aad9189. doi:10.1126/science.aad9189.

[6] Protopapa S, Grundy WM, Reuter DC, Hamilton DP, Dalle Ore CM, Cook JC, et al. Pluto's global surface composition through pixel-by-pixel Hapke modeling of New Horizons Ralph/LEISA data. Icarus 2017;287:218–228. doi:10.1016/j.icarus.2016.11.028.

[7] Cordiner MA, Nixon CA, Teanby NA, Irwin PGJ, Serigano J, Charnley SB, et al. ALMA Measurements of the HNC and $HC_3N$ Distributions in Titan's Atmosphere. The Astrophysical Journal Letters 2014;795:L30. doi:10.1088/2041-8205/795/2/L30.

[8] Lellouch E, Gurwell MA, Butler B, Fouchet T, Lavvas P, Strobel DF, et al. Detection of CO and HCN in Pluto's atmosphere with ALMA. Icarus 2017;286:289–307. doi:10.1016/j.icarus.2016.10.013.

[9] Brown LR, Humphrey CM, Gamache RR. $CO_2$-broadened water in the pure rotation and $v_2$ fundamental regions. Journal of Molecular Spectroscopy 2007;246:1. doi:10.1016/j.jms.2007.07.010.

[10] Harrison JJ, Bernath PF. Infrared absorption cross sections for propane (C3H8) in the 3μm region. Journal of Quantitative Spectroscopy and Radiative Transfer 2010;111:1282. doi:10.1016/j.jqsrt.2009.11.027.

[11] Quirico E, Schmitt B. A Spectroscopic Study of CO Diluted in N 2Ice: Applications for Triton and Pluto. Icarus 1997;128:181–188. doi:10.1006/icar.1997.5710.

[12] Edwards DP. GENLN2: A General Line-by-Line Atmospheric Transmittance and Radiance Model. The National Center for Atmospheric Research: Technical Note 367-STR; 1992.

[13] Clough SA, Shephard MW, Mlawer EJ, Delamere JS, Iacono MJ, Cady-Pereira K, et al. Atmospheric radiative transfer modeling: a summary of the AER codes. Journal of Quantitative Spectroscopy and Radiative Transfer 2005;91:233–244. doi:10.1016/j.jqsrt.2004.05.058.

[14] Gordon IE, Rothman LS, Hill C, Kochanov RV, Tan Y, Bernath PF, et al. The HITRAN2016 molecular spectroscopic database. Journal of Quantitative Spectroscopy and Radiative Transfer 2017;203:3–69. doi:10.1016/j.jqsrt.2017.06.038.

[15] Jacquinet-Husson N, Armante R, Scott NA, Chédin A, Crépeau L, Boutammine C, et al. The 2015 edition of the GEISA spectroscopic database. Journal of Molecular Spectroscopy 2016;327:31–72. doi:10.1016/j.jms.2016.06.007.

[16]    Pickett HM, Poynter RL, Cohen EA, DELITSKY ML, Pearson JC, Müller HSP. Submillimeter, millimeter, and microwave spectral line catalog. Journal of Quantitative Spectroscopy and Radiative Transfer 1998;60:883–890. doi:10.1016/S0022-4073(98)00091-0.

[17]    Müller HSP, Schlöder F, Stutzki J, Winnewisser G. The Cologne Database for Molecular Spectroscopy, CDMS: a useful tool for astronomers and spectroscopists. Journal of Molecular Structure 2005;742:215–227. doi:10.1016/j.molstruc.2005.01.027.

[18]    Alvarado MJ, Payne VH, Mlawer EJ, Uymin G, Shephard MW, Cady-Pereira KE, et al. Performance of the Line-By-Line Radiative Transfer Model (LBLRTM) for temperature, water vapor, and trace gas retrievals: recent updates evaluated with IASI case studies. Atmospheric Chemistry and Physics 2013;13:6687–6711. doi:10.5194/acp-13-6687-2013.

[19]    Millour E, Forget F, Spiga A, Navarro T, Madeleine JB, Montabone L, et al. The Mars Climate Database (MCD version 5.2). European Planetary Science Congress 2015 2015;10:EPSC2015–438.

[20]    Hanel RAC. Exploration of the solar system by infrared remote sensing. 1992.

[21]    Ehrenreich D, Tinetti G, Lecavelier des Etangs A, Vidal-Madjar A, Selsis F. The transmission spectrum of Earth-size transiting planets. Astronomy and Astrophysics 2006;448:379–393. doi:10.1051/0004-6361:20053861.

[22]    Moses JI, Fouchet T, Bézard B, Gladstone GR, Lellouch E, Feuchtgruber H. Photochemistry and diffusion in Jupiter's stratosphere: Constraints from ISO observations and comparisons with other giant planets. Journal of Geophysical Research 2005;110:8001. doi:10.1029/2005JE002411.

[23]    Marten A, Matthews HE, Owen T, Moreno R, Hidayat T, Biraud Y. Improved constraints on Neptune's atmosphere from submillimetre-wavelength observations. Astronomy and Astrophysics 2005;429:1097–1105. doi:10.1051/0004-6361:20041695.

[24]    Teanby NA, Irwin PGJ, de Kok R, Nixon CA, Coustenis A, Bézard B, et al. Latitudinal variations of HCN, HC 3N, and C 2N 2 in Titan's stratosphere derived from Cassini CIRS data. Icarus 2006;181:243–255. doi:10.1016/j.icarus.2005.11.008.

[25]    Curtis AR. Discussion of "A statistical model for water vapour absorption." Quarterly Journal of the Royal Meteorological Society 1952:638–640.

[26]    Godson WL. The evaluation of infra-red radiative fluxes due to atmospheric water vapour. Quarterly Journal of the Royal Meteorological Society 1953;79:367–379. doi:10.1002/qj.49707934104.

[27]    Robinson TD, Catling DC. Common 0.1 bar tropopause in thick atmospheres set by pressure-dependent infrared transparency. Nature Geoscience 2014;7:12–5. doi:10.1038/ngeo2020.

[28]    Villanueva GL, Mumma MJ, Bonev BP, Novak RE, Barber RJ, Disanti MA. Water in planetary and cometary atmospheres: $H_2O/HDO$ transmittance and fluorescence models. Journal of Quantitative Spectroscopy and Radiative Transfer 2012;113:202–220. doi:10.1016/j.jqsrt.2011.11.001.

[29]    Villanueva GL, Magee-Sauer K, Mumma MJ. Modeling of nitrogen compounds in cometary atmospheres: Fluorescence models of ammonia ($NH_3$), hydrogen cyanide (HCN), hydrogen isocyanide (HNC) and cyanoacetylene ($HC_3N$). Journal of Quantitative Spectroscopy and Radiative Transfer 2013;129:158–168. doi:10.1016/j.jqsrt.2013.06.010.

[30]    Villanueva GL, Mumma MJ, Magee-Sauer K. Ethane in planetary and cometary atmospheres: Transmittance and fluorescence models of the $\nu_7$ band at 3.3 μm. Journal of Geophysical Research 2011;116:1–23. doi:10.1029/2010JE003794.

[31]    Villanueva GL, Disanti MA, Mumma MJ, Xu L-H. A Quantum Band Model of the $\nu_3$ Fundamental of Methanol ($CH_3OH$) and its Application to Fluorescence Spectra of Comets. The Astrophysical Journal 2012;747:1–11. doi:10.1088/0004-637X/747/1/37.

[32]    Villanueva GL, Mumma MJ, Disanti MA, Bonev BP, Gibb EL, Magee-Sauer K, et al. The Molecular Composition of Comet C/2007 W1 (Boattini): Evidence of a Peculiar Outgassing and a Rich Chemistry. Icarus 2011;216:227–240. doi:10.1016/j.icarus.2011.08.024.

[33]    Kempton EMR, Lupu R, Owusu-Asare A, Slough P, Cale B. Exo-Transmit: An Open-Source Code for Calculating Transmission Spectra for Exoplanet Atmospheres of Varied Composition. Publications of the Astronomical Society of the Pacific 2017;129:044402–. doi:10.1088/1538-3873/aa61ef.

[34]    Keller-Rudek H, Moortgat GK, Sander R, Sörensen R. The MPI-Mainz UV/VIS Spectral Atlas of Gaseous Molecules of Atmospheric Interest. Earth System Science Data 2013;5:365–73. doi:https://doi.org/10.5194/essd-5-365-2013.

[35]    Serdyuchenko A, Gorshelev V, Weber M, Chehade W, Burrows JP. High spectral resolution ozone absorption cross-sections – Part 2: Temperature dependence. Atmos Meas Tech 2014;7:625–36. doi:10.5194/amt-7-625-2014.

[36]    Venot O, Bénilan Y, Fray N, Gazeau M-C, Lefèvre F, Es-sebbar E, et al. VUV-absorption cross section of carbon dioxide from 150 to 800 K and applications to warm exoplanetary atmospheres. A&A 2018;609:A34. doi:10.1051/0004-6361/201731295.

[37]    Gelaro R, McCarty W, Suarez MJ, Todling R, Molod A, Takacs L, et al. The Modern-Era Retrospective Analysis for Research and Applications, Version 2 (MERRA-2). Journal of Climate 2017;30:5419–5454. doi:10.1175/JCLI-D-16-0758.1.

[38]    Gesch DB, Larson KS. Techniques for development of global 1-kilometer digital elevation models. Pecora Thirteen, Human Interactions with the Environment - Perspectives from Space, Sioux Falls, South Dakota: 1996.

[39]    Smith DE, Zuber MT, Frey HV, Garvin JB, Head JW, Muhleman DO, et al. Mars Orbiter Laser Altimeter: Experiment summary after the first year of global mapping of Mars. Journal of Geophysical Research 2001;106:23689. doi:10.1029/2000JE001364.

[40]    Smith MD. Interannual variability in TES atmospheric observations of Mars during 1999-2003. Icarus 2004;167:148–165. doi:10.1016/j.icarus.2003.09.010.

[41]    Sánchez-Lega A, Heimpel M. Atmospheric Dynamics of Giants and Icy Planets. Handbook of Exoplanets, Springer, Cham; 2017, p. 1–32. doi:10.1007/978-3-319-30648-3_51-1.

[42]    Knollenberg R., Travis L., Tomasko M., Smith P., Ragent B., Esposito L., et al. The clouds of Venus: A synthesis report. Journal of Geophysical Research: Space Physics 1980;85:8059–81. doi:10.1029/JA085iA13p08059.

[43]    Lellouch E, Hartogh P, Feuchtgruber H, Vandenbussche B, Graauw T de, Moreno R, et al. First results of Herschel-PACS observations of Neptune. A&A 2010;518:L152. doi:10.1051/0004-6361/201014600.

[44]    Ehrenreich D, Vidal-Madjar A, Widemann T, Gronoff G, Tanga P, Barthélemy M, et al. Transmission spectrum of Venus as a transiting exoplanet. Astronomy and Astrophysics 2012;537:L2. doi:10.1051/0004-6361/201118400.

[45]    Devi VM, Benner DC, Sung K, Crawford TJ, Gamache RR, Renaud CL, et al. Line parameters for CO2 broadening in the v2 band of HD16O. Journal of Quantitative Spectroscopy and Radiative Transfer 2017;187:472–88. doi:10.1016/j.jqsrt.2016.10.004.

[46]    Devi VM, Benner DC, Sung K, Crawford TJ, Gamache RR, Renaud CL, et al. Line parameters for CO2- and self-broadening in the v1 band of HD16O. Journal of Quantitative Spectroscopy and Radiative Transfer 2017;203:133–57. doi:10.1016/j.jqsrt.2017.01.032.

[47]    Gamache RR, Renaud CL. Modified Complex Robert-Bonamy (MCRB) calculations of H 2 O transitions broadened by H 2 for applications to planetary and exoplanet atmospheres. J Phys: Conf Ser 2017;810:012032. doi:10.1088/1742-6596/810/1/012032.

[48]    Delsemme AH. Chemical composition of cometary nuclei, 1982, p. 85–130.

[49]    Biver N, Bockelée-Morvan D, Colom P, Crovisier J, Henry F, Lellouch E, et al. The 1995 2002 Long-Term Monitoring of Comet C/1995 O1 (HALE BOPP) at Radio Wavelength. Earth 2002;90:5–14. doi:10.1023/A:1021599915018.

[50]    Jorda L, Crovisier J, Green DWE. The Correlation Between Visual Magnitudes and Water Production Rates. Asteroids, Comet, Meteors 2008 2008:8046.

[51]    Mumma MJ, Disanti MA, Dello Russo N, Magee-Sauer K, Gibb EL, Novak R. Remote infrared observations of parent volatiles in comets: A window on the early solar system. Advances in Space Research 2003;31:2563–2575. doi:10.1016/S0273-1177(03)00578-7.

[52]    Crovisier J. Cometary diversity and cometary families. ArXiv Astrophysics E-Prints 2007:arXiv:astro-ph/0703785.

[53]    Villanueva GL, Bonev BP, Mumma MJ, Magee-Sauer K, Disanti MA, Salyk C, et al. The Volatile Composition of the Split Ecliptic comet 73P/Schwassmann-Wachmann 3: A Comparison of Fragments C and B. The Astrophysical Journal 2006;650:L87–L90. doi:10.1086/508806.

[54]    Mumma MJ, Charnley SB. The Chemical Composition of Comets – Emerging Taxonomies & Natal Heritage. Annual Review of Astronomy and Astrophysics 2011;49:471–524.

[55]    Parmentier V, Guillot T. A non-grey analytical model for irradiated atmospheres. I. Derivation. Astronomy and Astrophysics 2014;562:A133. doi:10.1051/0004-6361/201322342.

[56]    Pieters CM, Goswami JN, Clark RN, Annadurai M, Boardman J, Buratti B, et al. Character and Spatial Distribution of OH/H2O on the Surface of the Moon Seen by M3 on Chandrayaan-1. Science 2009;326:568–. doi:10.1126/science.1178658.

[57]    Combe J-P, McCord TB, McFadden LA, Ieva S, Tosi F, Longobardo A, et al. Composition of the northern regions of Vesta analyzed by the Dawn mission. Icarus 2015;259:53–71. doi:10.1016/j.icarus.2015.04.026.

[58]    Protopapa S, Böhnhardt H, Herbst TM, Cruikshank DP, Grundy WM, Merlin F, et al. Surface characterization of Pluto and Charon by L and M band spectra. Astronomy and Astrophysics 2008;490:365–375. doi:10.1051/0004-6361:200809994.

[59]    Protopapa S, Sunshine JM, Feaga LM, Kelley MSP, A'Hearn MF, Farnham TL, et al. Water ice and dust in the innermost coma of comet 103P/Hartley 2. Icarus 2014;238:191–204. doi:10.1016/j.icarus.2014.04.008.

[60]    Hapke B. Bidirectional reflectance spectroscopy. 1. Theory. Journal of Geophysical Research 1981;86:4571–4586.

[61]    Hapke B. Theory of reflectance and emittance spectroscopy. Topics in Remote Sensing 1993.

[62]   Viviano-Beck CE, Seelos FP, Murchie SL, Kahn EG, Seelos KD, Taylor HW, et al. Revised CRISM spectral parameters and summary products based on the currently detected mineral diversity on Mars. Journal of Geophysical Research 2014;119:1403–1431. doi:10.1002/2014JE004627.

[63]   Clark RN, Swayze GA, Wise, R, Livo E, Hoefen TM, Kokaly R, et al. USGS 2007 Spectral Library splib06a, Data Series 231. U.S. Geological Survey; 2007.

[64]   Jäger C, B. Il'in V, Henning T, Mutschke H, Fabian D, A. Semenov D, et al. A database of optical constants of cosmic dust analogs. Journal of Quantitative Spectroscopy and Radiative Transfer 2003;79–80:765–74. doi:10.1016/S0022-4073(02)00301-1.

[65]   Hudson RL, Gerakines PA, Moore MH. Infrared spectra and optical constants of astronomical ices: II. Ethane and ethylene. Icarus 2014;243:148–57. doi:10.1016/j.icarus.2014.09.001.

[66]   Bauschlicher CW, Ricca A, Boersma C, Allamandola LJ. The NASA Ames PAH IR Spectroscopic Database: Computational Version 3.00 with Updated Content and the Introduction of Multiple Scaling Factors. ApJS 2018;234:32. doi:10.3847/1538-4365/aaa019.

[67]   Massie ST, Hervig M. HITRAN 2012 refractive indices. Journal of Quantitative Spectroscopy and Radiative Transfer 2013;130:373–380. doi:10.1016/j.jqsrt.2013.06.022.

[68]   DeMeo FE, Binzel RP, Slivan SM, Bus SJ. An extension of the Bus asteroid taxonomy into the near-infrared. Icarus 2009;202:160–180. doi:10.1016/j.icarus.2009.02.005.

[69]   Smith MD, Wolff MJ, Clancy RT, Kleinböhl A, Murchie SL. Vertical distribution of dust and water ice aerosols from CRISM limb-geometry observations. Journal of Geophysical Research 2013;118:321–334. doi:10.1002/jgre.20047.

[70]   Goldsmith PF, Langer WD. Population Diagram Analysis of Molecular Line Emission. The Astrophysical Journal 1999;517:209–225. doi:10.1086/307195.

[71]   Kurucz RL. Including All the Lines. AIP Conference Proceedings, Harvard-Smithsonian Center for Astrophysics, Cambridge, MA 02138, USA; 2009, p. 43–51. doi:10.1063/1.3250087.

[72]   Hase F, Wallace L, Mcleod SD, Harrison JJ, Bernath PF. The ACE-FTS atlas of the infrared solar spectrum. Journal of Quantitative Spectroscopy and Radiative Transfer 2009;111:521–528. doi:10.1016/j.jqsrt.2009.10.020.

[73]   Humlícek J. Optimized computation of the Voigt and complex probability functions. Journal of Quantitative Spectroscopy and Radiative Transfer 1982;27:437–444. doi:10.1016/0022-4073(82)90078-4.

[74]   Wells RJ. Rapid approximation to the Voigt/Faddeeva function and its derivatives. Journal of Quantitative Spectroscopy and Radiative Transfer 1999;62:29–48. doi:10.1016/S0022-4073(97)00231-8.

[75]   Gordley LL, Marshall BT, Chu DA. Linepak: Algorithms for modeling spectral transmittance and radiance. Journal of Quantitative Spectroscopy & Radiative Transfer (ISSN 0022-4073) 1994;52:563–580. doi:10.1016/0022-4073(94)90025-6.

[76]   Chou MD, Arking A. Computation of Infrared Cooling Rates in the Water Vapor Bands. Journal of Atmospheric Sciences 1980;37:855–867. doi:10.1175/1520-0469(1980)037<0855:COICRI>2.0.CO;2.

[77]   Stamnes K, Tsay SC, Jayaweera K, Wiscombe W. Numerically stable algorithm for discrete-ordinate-method radiative transfer in multiple scattering and emitting layered media. Applied Optics 1988;27:2502–2509. doi:10.1364/AO.27.002502.

[78]    Lin Z, Stamnes S, Jin Z, Laszlo I, Tsay SC, Wiscombe WJ, et al. Improved discrete ordinate solutions in the presence of an anisotropically reflecting lower boundary: Upgrades of the DISORT computational tool. Journal of Quantitative Spectroscopy and Radiative Transfer 2015;157:119–134. doi:10.1016/j.jqsrt.2015.02.014.

[79]    Turner DD. Arctic Mixed-Phase Cloud Properties from AERI Lidar Observations: Algorithm and Results from SHEBA. Journal of Applied Meteorology 2005;44:427–444. doi:10.1175/JAM2208.1.

[80]    Goody RM, Yung YL. Atmospheric Radiation, A Theoretical Basis. Oxford Universtiy Press; 1995.

[81]    Thomas GE, Stamnes K. Radiative Transfer in the Atmosphere and Ocean. Cambridge University Press; 2002.

[82]    Wiscombe WJ. Improved Mie scattering algorithms. Applied Optics 1980;19:1505–9. doi:10.1364/AO.19.001505.

[83]    Mishchenko MI, Travis LD, Mackowski DW. T-matrix computations of light scattering by nonspherical particles: a review. Journal of Quantitative Spectroscopy and Radiative Transfer 1996;55:535–75. doi:10.1016/0022-4073(96)00002-7.

[84]    Wolff MJ, Smith MD, Clancy RT, Arvidson R, Kahre M, Seelos F, et al. Wavelength dependence of dust aerosol single scattering albedo as observed by the Compact Reconnaissance Imaging Spectrometer. Journal of Geophysical Research 2009;114:E00D04. doi:10.1029/2009JE003350.

[85]    Bohren CF, Huffman DR. Absorption and scattering of light by small particles. 1983.

[86]    Spurr R. Simultaneous derivation of intensities and weighting functions in a general pseudo-spherical discrete ordinate radiative transfer treatment. Journal of Quantitative Spectroscopy and Radiative Transfer 2002;75:129–75. doi:10.1016/S0022-4073(01)00245-X.

[87]    Xie X, Mumma MJ. Monte Carlo Simulation of Cometary Atmospheres: Application to Comet P/Halley at the Time of the Giotto Spacecraft Encounter. II. Axisymmetric Model. The Astrophysical Journal 1996;464:457–475. doi:10.1086/177336.

[88]    Leinert C, Bowyer S, Haikala LK, Hanner MS, Hauser MG, Levasseur-Regourd A-C, et al. The 1997 reference of diffuse night sky brightness. Astronomy and Astrophysics Supplement Series 1998;127:1–99. doi:10.1051/aas:1998105.

[89]    Zmuidzinas J. Thermal noise and correlations in photon detection. Applied Optics 2003;42:4989–5008. doi:10.1364/AO.42.004989.

[90]    Bridger A, Clarke D, Lorente NPF, Yatagai H, Schilling M, Testi L, et al. ALMA Observing Tool. vol. 7019, 2008, p. 70190R. doi:10.1117/12.789108.

[91]    Markwardt CB. Non-linear Least-squares Fitting in IDL with MPFIT. vol. 411, Eprint: ArXiv:0902.2850: 2009, p. 251.

[92]    Villanueva GL, Mumma MJ, Novak RE, Radeva YL, Käufl HU, Smette A, et al. A sensitive search for organics ($CH_4$, $CH_3OH$, $H_2CO$, $C_2H_6$, $C_2H_2$, $C_2H_4$), hydroperoxyl ($HO_2$), nitrogen compounds ($N_2O$, $NH_3$, $HCN$) and chlorine species ($HCl$, $CH_3Cl$) on Mars using ground-based high-resolution infrared spectroscopy. Icarus 2013;223:11–27. doi:10.1016/j.icarus.2012.11.013.

[93]    Yurchenko S, Al-Refaie A, Tennyson J. ExoCross: a general program for generating spectra from molecular line lists. Astronomy & Astrophysics 2018. doi:10.1051/0004-6361/201732531.

[94]    Encrenaz T, de Graauw T, Schaeidt S, Lellouch E, Feuchtgruber H, Beintema DA, et al. First results of ISO-SWS observations of Jupiter. Astronomy and Astrophysics 1996;315:L397–L400.

[95]    Marten A, Matthews HE, Owen T, Moreno R, Hidayat T, Biraud Y. Improved constraints on Neptune's atmosphere from submillimetre-wavelength observations. Astronomy and Astrophysics 2005;429:1097–1105. doi:10.1051/0004-6361:20041695.

[96]    Bétrémieux Y, Kaltenegger L. Transmission Spectrum of Earth as a Transiting Exoplanet from the Ultraviolet to the Near-infrared. The Astrophysical Journal Letters 2013;772:L31. doi:10.1088/2041-8205/772/2/L31.

[97]    Arney GN, Meadows VS, Domagal-Goldman SD, Deming D, Robinson TD, Tovar G, et al. Pale Orange Dots: The Impact of Organic Haze on the Habitability and Detectability of Earthlike Exoplanets. The Astrophysical Journal 2017;836:49. doi:10.3847/1538-4357/836/1/49.